\documentclass{aa} 
\usepackage{graphicx} 
\usepackage{txfonts} 
\usepackage{natbib}	 
\usepackage{ulem} 
\usepackage{txfonts,epsfig,graphicx,url,twoopt} 
\usepackage[breaklinks=true]{hyperref} 
\usepackage{dblfloatfix} 
\usepackage{capt-of} 
\hypersetup{colorlinks=true,citecolor=blue} 
\bibpunct{(}{)}{;}{a}{}{,} 
\begin{document}

   \title{Simulated Galactic methanol maser distribution to constrain
Milky Way parameters}                                                   
   \author{L.~H.~Quiroga-Nu\~{n}ez 
   \inst{1,2}, 
    Huib~Jan~van~Langevelde 
    \inst{2,1}, 
     M.~J.~Reid 
     \inst{3} 
     \and 
     J.~A.~Green 
     \inst{4,5} 
     } 
 
    \institute{Leiden Observatory, Leiden University, P.O. Box 9513,
2300 RA Leiden, The Netherlands.\\                                      
              \email{quiroganunez@strw.leidenuniv.nl} 
        	       \and 
	      Joint Institute for VLBI ERIC (JIVE), Postbus 2, 7990 AA Dwingeloo,
The Netherlands.                                                        
              \and 
	      Harvard-Smithsonian Center for Astrophysics, 60 Garden Street,
Cambridge, MA 02138, USA.                                               
              \and 
	      CSIRO Astronomy and Space Science, Australia Telescope National
Facility, PO Box 76, Epping, NSW 1710, Australia.                       
	     \and 
	      SKA Organisation, Jodrell Bank Observatory, Lower Withington,
Macclesfield, SK11 9DL, UK.                                             
	      } 
 
   \date{Received 23 February 2017 / Accepted 7 June 2017}

 
  \abstract 
{Using trigonometric parallaxes and proper motions of masers associated
with massive young stars, the Bar and Spiral Structure Legacy (BeSSeL)
survey has reported the most accurate values of the Galactic parameters
so far. The determination of these parameters with high accuracy
has a widespread impact on Galactic and extragalactic measurements.}    
{This research is aimed at establishing the confidence with which
such parameters can be determined. This is relevant for the data
published in the context of the BeSSeL survey collaboration, but
also for future observations, in particular from the southern hemisphere.
In addition, some astrophysical properties of the masers can be
constrained, notably the luminosity function.}                          
   {We have simulated the population of maser-bearing young stars
associated with Galactic spiral structure, generating several
samples and comparing them with the observed samples used in the
BeSSeL survey. Consequently, we checked the determination of Galactic
parameters for observational biases introduced by the sample selection.}
{Galactic parameters obtained by the BeSSeL survey do not seem to
be biased by the sample selection used. In fact, the published error
estimates appear to be conservative for most of the parameters. We
show that future BeSSeL data and future observations with southern
arrays will improve the Galactic parameters estimates and smoothly
reduce their mutual correlation. Moreover, by modeling future parallax
data with larger distance values and, thus, greater relative uncertainties
for a larger numbers of sources, we found that parallax-distance
biasing is an important issue. Hence, using fractional parallax uncertainty
in the weighting of the motion data is imperative. Finally, the luminosity
function for 6.7 GHz methanol masers was determined, allowing us
to estimate the number of Galactic methanol masers.}                    
   {} 
 
\authorrunning{L.H. Quiroga-Nu\~{n}ez et al.} 
\titlerunning{Simulated maser distribution to constrain Milky Way
parameters}                                                             
\maketitle 
 
 
\section{Introduction} 
 
A lack of accurate distance measurements throughout the Galaxy combined
with our location within the Milky Way have complicated the interpretation
of astrometric measurements~\citep{2014ARA&A..52..339R}. Consequently,
the most fundamental Galactic parameters, such as the distance to
the Galactic center ($R_0$), the rotation speed at the solar radius
($\Theta_0 $), and the rotation curve (e.g., $d\Theta/dR$) have not been
established with high accuracy. At Galactic scales, distance estimates
through radial velocities, mass and luminosity calculations of sources
within the Galaxy, as well as the mass and luminosity estimates of
the Milky Way, depend on the Galactic parameters. Additionally, extragalactic
measurements are based on Galactic calibrations that are made using
the Milky Way parameter values. Therefore, highly accurate estimates
of the fundamental Galactic parameters are vitally important.           
 
A step forward came with the Hipparcos satellite~\citep{1997A&A...323L..49P}.
It provided astrometric accuracies of the order of 1 milliarcsecond
(mas), which allows distance estimations in the solar neighborhood
($\sim$100 pc) with 10$\%$ accuracy. However, this is a tiny portion
of the Milky Way. The ongoing European Space Agency mission, Gaia,
aims to measure parallaxes and proper motions of $10^9$ stars with
accuracies up to 20 $\mu$as at 15 mag with a distance horizon of
5 kpc with $10\%$ accuracy and 10 kpc with $20\%$ accuracy~\citep{2001A&A...369..339P,2016A&A...595A...2G}.
Although Gaia will transform our knowledge of the Milky Way, the
mission is restricted to optical wavelengths and due to significant
dust obscuration, it will not be able to probe the Galactic plane
freely. In contrast, radio wavelengths are not affected by dust extinction
and can be used throughout the Galaxy.                                  
 
Direct accurate distances and proper motions have been measured for
maser-bearing young stars~\citep[e.g.][]{2014ApJ...781..108S,2017MNRAS.tmp..217B};
this data was obtained employing Very Long Baseline Interferometry
(VLBI). This astrometric information has provided us with a better
understanding of the Milky Way's spiral structure, insights into the
formation and evolution of our Galaxy, its 3D gravitational
potential, and the Galactic baryonic and dark matter distribution~\citep{2011ARep...55..108E}.
 
The most suitable radio beacons for astrometry are methanol (6.7
and 12.2 GHz) and water (22 GHz) masers~\citep{2011AN....332..461B}.
In addition to being bright, water masers can be associated with high mass
star forming regions (HMSFRs), while class II 6.7 and 12 GHz methanol
masers are uniquely associated with HMSFRs~\citep[e.g.][]{2013MNRAS.435..524B,2013A&A...556A..73S}.
By detecting 6.7 GHz methanol masers, we trace the Galactic
spiral structure because HMSFRs are expected to be born close to
a spiral arm and evolve more quickly than low-mass stars~\citep{2013MNRAS.433.1114Y}.
Therefore, HMSFRs should follow the disk rotation with low dispersion
(compared, for example, to masers in evolved stars).                   
 
Given parallax, proper motion measurements, source coordinates, and
line-of-sight velocities (from Doppler shifts of spectral lines)
to methanol and water masers, it is possible to sample complete phase-space
information. This provides direct and powerful constraints on the
fundamental parameters of the Galaxy. The Bar and Spiral Structure
Legacy (BeSSeL\footnote{\url{http://bessel.vlbi-astrometry.org/}})
survey has addressed this task using different arrays: the Very Long
Baseline Array (VLBA) in USA and the European VLBI Network (EVN)
in Europe, Asia and South Africa.  Additionally similar parallax and proper motion 
data has come from the VLBI Exploration of Radio Astrometry (VERA) in Japan.
The most recent summary paper \citep{2014ApJ...783..130R} lists
astrometric data for 103 parallax measurements with typical accuracies
of $20 \mu$as. By fitting these sources to an axially symmetric Galactic
model, they provide accurate values for the fundamental
Galactic parameters: $R_0 = 8.34 \pm 0.16$ kpc, $\Theta_0 = 240 \pm
8 \ \rm{km \, s^{-1}}$, and $d\Theta/dR=-0.2 \pm 0.4 \rm{km \, s^{-1}
\, kpc^{-1}}$ between Galactocentric radii of 5 and 16 kpc.
 
Although the BeSSeL survey data is very accurate, the target selection
used was necessarily biased. It has targeted the brightest known
masers accessible to the (northern hemisphere) VLBI arrays used.
Most of the published targets used by BeSSeL for astrometric measurements
are 22 GHz water masers and 12 GHz methanol masers that were originally
selected based on 6.7 GHz surveys. In the current study,
a model used to simulate the 6.7 GHz methanol maser distribution in the
Milky Way is presented. The model was compared with systematic surveys,
allowing us to determine the luminosity function. Also, it is used
to generate different artificial samples that can be used to test how accurately
they can fit a Galactic model and how a given
level of incompleteness can bias the Galactic parameter values. This
is particularly important when more sources are being added
to the BeSSeL sample.                                                  
 
In Sect.~\ref{model} the components and assumptions of the model
are presented. Next, Sect.~\ref{results} describes the luminosity
function fitted using observational surveys, the Galactic parameter
results, and the correlation among parameters using several samples.
Finally, the discussion and conclusions of the results compared to
the BeSSeL findings are shown in Sects.~\ref{discussion} and~\ref{conclusions},
respectively.                                                           
 

\section{Model for the 6.7 GHz methanol maser distribution in the
spiral structure}                                                       
\label{model} 
 
The main components of the Milky Way can be identified as a halo,
nuclear bulge (or bar), and two disk components: a thin and a thick disk~\citep[see
e.g.][]{1983MNRAS.202.1025G,2013A&ARv..21...61R}. The current model
is centered on the thin disk component, more specifically on a 
spiral structure between 3 kpc and 15 kpc as traced by HMSFRs that 
contains methanol maser bearing stars. 
Following the analysis made by~\cite{2014ApJ...783..130R}, the model is based on a galaxy
with spiral structure. The analysis of the rotation and scale of the galaxy does not seem directly dependent on this assumption.                           
 
\begin{table} 
\centering 
\begin{tabular}{c c c} 
\hline\hline 
Model & Distribution & Distribution\\ 
Variable & Type & Parameters \\ 
\hline 
Galactic&Radial decay and &$h_r=2.44$ kpc (1)\\ 
Plane (X,Y)&Monte Carlo rejection&$\sigma_d=$ 0.35 kpc (1)\\ 
\hline 
Vertical&Gaussian&$\mu_z=0$ kpc\\ 
Position (Z)  (2) & &$\sigma_z=25$ pc\\ 
\hline 
Radial&Gaussian&$\mu_r=0$ km $\rm{s^{-1}}$\\ 
Velocity (U)& &$\sigma_r=5$ km $\rm{s^{-1}}$\\ 
\hline 
Tangential&Gaussian&$\mu_t=\Theta_0=240$ km $\rm{s^{-1}}$ (1)\\ 
Velocity (V)& &$\sigma_t=9$ km $\rm{s^{-1}}$\\ 
\hline 
Vertical&Gaussian&$\mu_v=0$ km $\rm{s^{-1}}$\\ 
Velocity (W)& &$\sigma_v=5$ km $\rm{s^{-1}}$\\ 
\hline 
Luminosity&Power&Cutoffs: $10^{-8} \, \rm{L_{\sun}}$, $10^{-3} \,
\rm{L_{\sun}}$\\                                                        
Function (L) (3) &Law&and $\alpha=-1.43$ \\ 
\hline 
\end{tabular} 
\caption{\label{model_sum} Spatial, velocity and luminosity distributions
used in the current model.  We assumed the sun's vertical position to be z=0 pc and
any change in this value, which~\cite{2014ApJ...797...53G} suggested
to be 25 pc, was found negligible at these scales. References: (1)~\cite{2014ApJ...783..130R},
(2)~\cite{2011MNRAS.417.2500G}, and (3)~\cite{2007A&A...463.1009P}. Radial, tangential
and vertical velocity dispersion values are discussed in Sect.~\ref{vd}.}    
\end{table} 
 
The aim of the model is to build a simulated database ready to be
processed with the Galactic parameter fitting method used by the
BeSSeL survey. To do this, each simulated 6.7 GHz methanol maser has
spatial coordinates, velocity components, and an associated intrinsic
luminosity (and their respective uncertainties). In the following subsections,
we explain each of the distributions and the initial parameters adopted,
as well as the fitting procedure used to obtain the Galactic parameters
from the astrometric data. Table~\ref{model_sum} presents a summary
of the distributions and values used.                                   
 
\subsection{Initial parameters} 
\label{ip} 
 
\cite{2014ApJ...783..130R} presented their best estimates
of the Galactic parameter values (Model A5), which we adopt here
(see Table~\ref{initial}):                                            
 
\begin{figure*}%
  \resizebox{0.57\hsize}{!}{\includegraphics{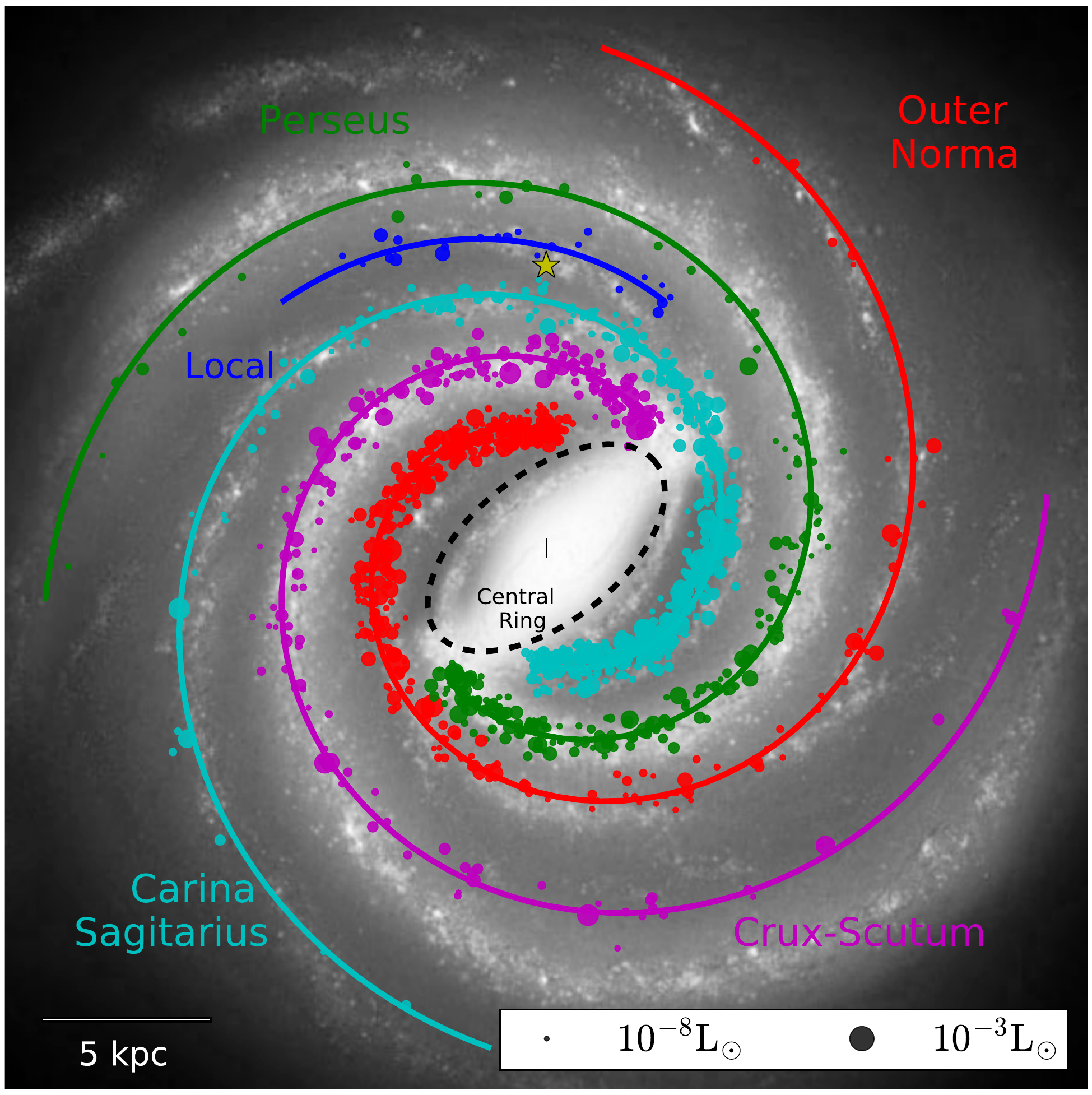}} 
   \resizebox{0.43\hsize}{!}{\includegraphics{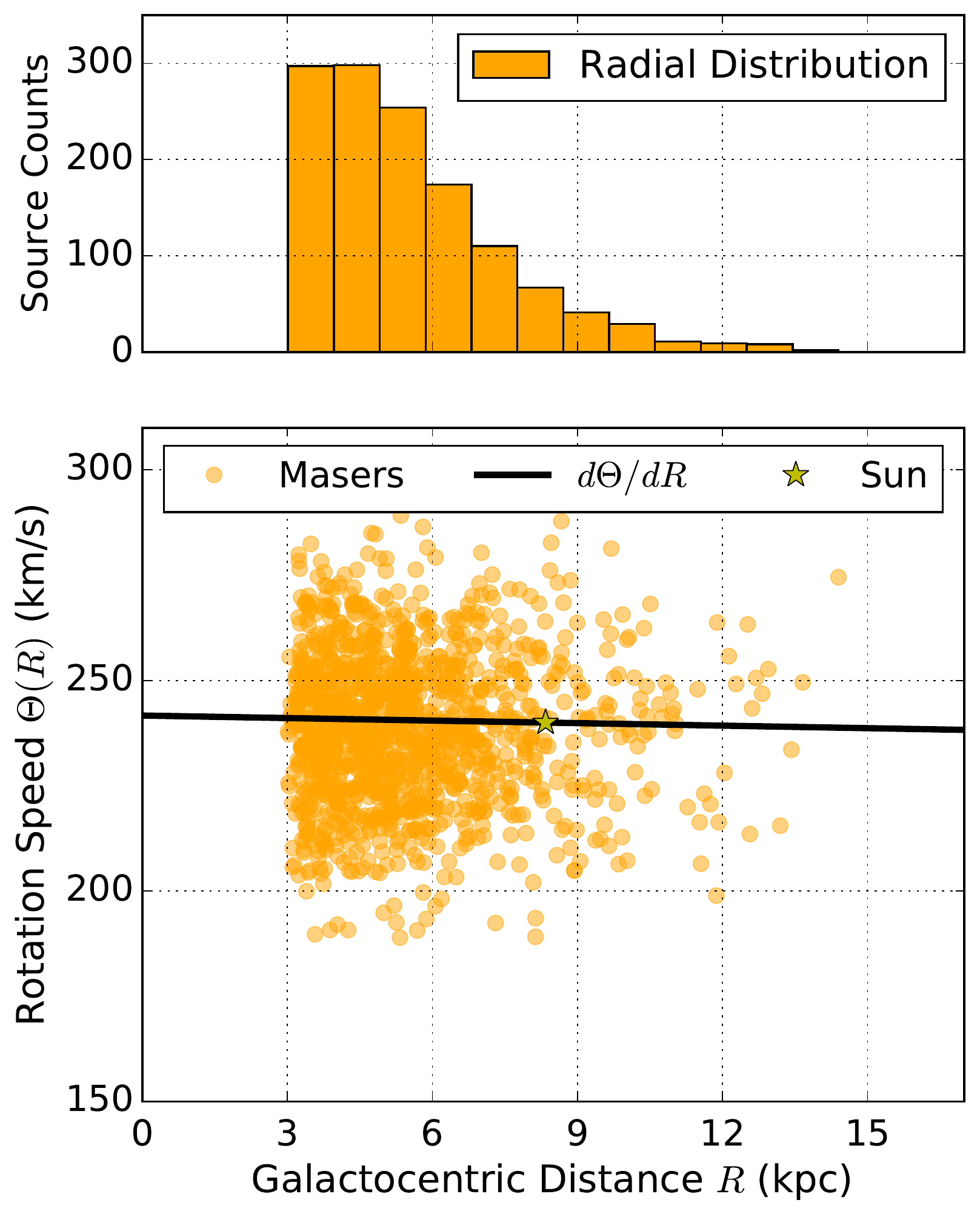}} 
  \caption{\label{samplot} \textbf{Left:} Galactic plane distribution
of 6.7 GHz methanol masers seen from the NGP overlaid on an artist
impression of the Milky Way (R. Hurt: NASA/JPLCaltech/SSC). The spiral
structure was constructed following~\cite{1992ApJS...83..111W} and
the central molecular ring or 3 kpc arms~\citep{2011ApJ...733...27G}
was indicated, but it is not part of the model. The simulated spatial
maser distribution is presented in Sect.~\ref{sd}. The plot also
includes the intrinsic peak luminosity for each source as the point
size following the luminosity function described in Sect.~\ref{ld}.
In this figure, the Galaxy rotates clockwise. \textbf{Bottom right:}
Tangential velocity distribution as a function of Galactocentric
distance for the simulated 6.7 GHz methanol masers. It also displays
the rotation curve, $d\Theta/dR = -0.2$ $\rm{km \, s^{-1} \, kpc^{-1}}$.
\textbf{Top right:} Radial distribution of the 6.7 GHz methanol
masers for our model is also shown. Table~\ref{model_sum} presents
a summary of the distributions used.}                                   
\end{figure*}%
 
\begin{itemize} 
\item $R_0,\Theta_0,d\Theta/dR$: \textbf{fundamental Galactic parameters.}
We took the current results of the BeSSeL survey, which assumes 
a Galactic model as a disk rotating at a speed of $\Theta(R) = \Theta_0
+ \frac{d\Theta}{dR} (R-R_0)$;                                          
\item $\bar{U}_s,\bar{V}_s$: \textbf{average source peculiar motion.}
When velocities are measured, systematic extra velocity components
can appear as a result of two effects: gas approaching a spiral arm
with enhanced gravitational attraction and magneto-hydrodynamic shocks
as the gas enters the arm; therefore, these extra velocity components,
which are defined at the position of each source, account for any average
peculiar motion of the masers;                                         
\item $U_{\sun},V_{\sun},W_{\sun}$: \textbf{solar motion.} Because
the model predicts the velocities with respect to the local
standard of rest (LSR) for all masers, the solar motion must be taken
into account in order to make the proper heliocentric corrections; 
\item $N$: \textbf{number of sources.} The total number of 6.7 GHz
methanol masers in the Galaxy is a required parameter to populate
the spiral arms. In Sect.~\ref{results}, this parameter is fitted
by comparing the model with the results of Methanol Multibeam Survey~\citep[MMB,
see:][]{2009MNRAS.392..783G,2010MNRAS.409..913G,2012MNRAS.420.3108G,2010MNRAS.404.1029C,2011MNRAS.417.1964C}
results given the adopted spatial distribution (Sect.~\ref{sd}) and luminosity
function (Sect.~\ref{ld}).                                     
\end{itemize} 
 
\begin{table} 
\centering 
\resizebox{\hsize}{!}{ 
\begin{tabular}{c c c} 
\hline \hline 
Parameter & Definition & Value \\ 
\hline 
$R_0$&Sun-Galactocentric distance&8.34 \ kpc \\ 
$\Theta_0$&Solar rotation speed&240 km $\rm{s^{-1}}$ \\ 
$d\Theta/dR$&Rotation curve&-0.2 km $\rm{s^{-1} \ kpc^{-1}}$ \\ 
$U_{\sun}$& Inward radial solar velocity&10.7 km $\rm{s^{-1}}$ \\ 
$V_{\sun}$&Tangential solar velocity&15.6 km $\rm{s^{-1}}$ \\ 
$W_{\sun}$&Vertical solar velocity&8.9 km $\rm{s^{-1}}$ \\ 
$\bar{U}_s$&Inward radial average peculiar motion&2.9 km $\rm{s^{-1}}$
\\                                                                      
$\bar{V}_s$&Tangential average peculiar motion& -1.5 km $\rm{s^{-1}}$
\\                                                                      
\hline 
\end{tabular}} 
\caption{\label{initial} Description of initial parameters values
used in the model which are based on the Model A5 results published
in~\cite{2014ApJ...783..130R}.}                                         
\end{table} 
 
\subsection{Spatial distribution} 
\label{sd} 
 
\begin{figure*}%
  \centering 
  \resizebox{\hsize}{!}{\includegraphics{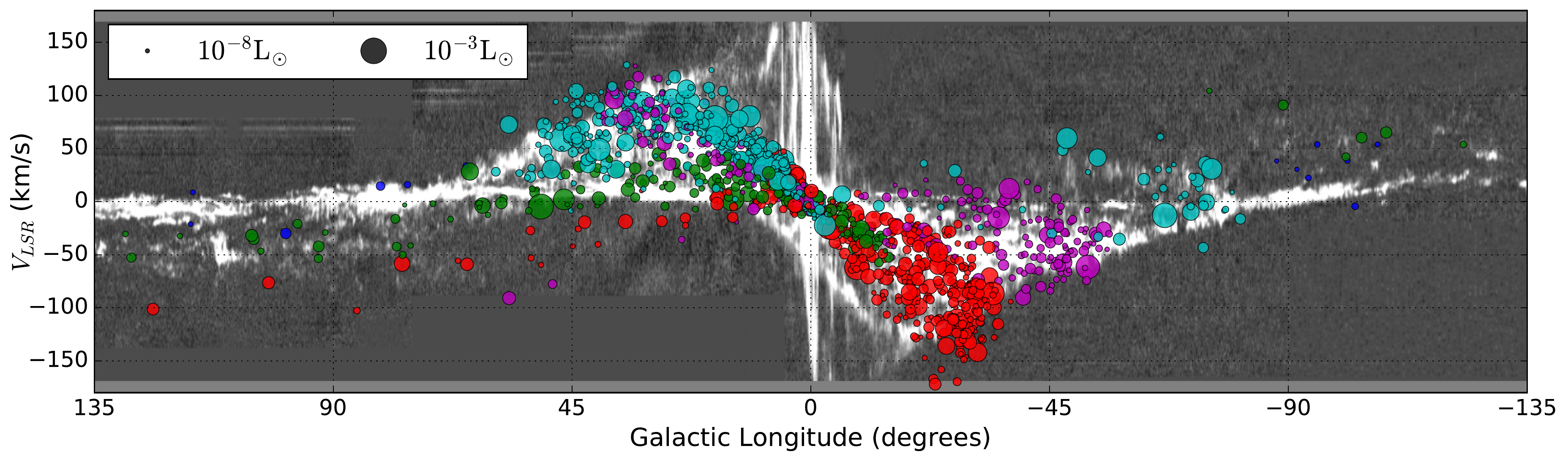}} 
  \caption{\label{lv} Velocity with respect to the LSR as a function of the
Galactic longitude for the simulated 6.7 GHz methanol masers distribution.
The point size is a measure of the peak luminosity function (Sect.~\ref{ld}).
Masers associated with different spiral
arms are color-coded as in Fig.~\ref{samplot}. The figure
is overlaid on the CO emission ($J =1-0$) plotted in grayscale and
taken from~\cite{2001ApJ...547..792D}.}                                 
\end{figure*}%
 
The spatial distribution along the spiral arms can be split into
two components, a Galactic plane distribution and a vertical component
distribution (Z). The latter can be drawn using a random generator
from a Gaussian distribution with a mean of 0 pc and $\sigma=$ 25
pc since massive young stars are found to be born close to the Galactic
plane~\citep[see e.g.][]{2011MNRAS.417.2500G,2016AstL...42..182B}.      
 
The Galactic plane distribution is drawn following two constraints.
First, the density of HMSFRs falls off exponentially with the Galactocentric
distance ($R$)~\citep{2013ApJ...779..115B}, and second, each source
should be associated with a spiral arm~\citep{2014ApJ...783..130R}.
For the first constraint, the maser radial distribution follows~\citet{2012ApJ...752...51C}
 
\begin{equation}\label{exp} 
n(R) \propto e^{-R/h_{R}}, 
\end{equation} 
 
\noindent where $n(R)$ is the number of sources and $h_{R}$ the exponential
scale length, which has been estimated from the maser parallax data
assuming a Persic Universal
rotation curve formulation to be $2.44$ kpc~\citep{2014ApJ...783..130R}
which we assumed valid for massive young stars. The top right panel of 
Fig.~\ref{samplot} shows the radial distribution of the simulated
masers. 

For the second constraint, the spiral arm positions
were set following an analytic approximation made by~\cite{1992ApJS...83..111W}.
Each spiral arm (four main arms and the local arm) can be located in
the Galactic plane using a simple relation in polar coordinates.
The left plot of Fig.~\ref{samplot} depicts the position of the
spiral arms as seen from the north galactic pole (NGP).                    
 In order to populate the spiral arms with 6.7 GHz methanol masers,
a rejection sampling Monte Carlo method was implemented. For this,
the model takes a source from the radial distribution (Eq.~\ref{exp})
and then the distance is calculated between the source and the closest
spiral arm. That distance $d$ is evaluated in a probability density
function of a Gaussian distribution

\begin{equation}\label{mu} 
P(d) \propto 
 \rm{exp} \, \left (\frac{-(d-\mu)^2}{2\sigma_d^2} \right ),\ 
\end{equation} 
 
\noindent where $\mu=0$ kpc, yielding the same likelihood of the
source to be behind or in front of the spiral arm. We took $\sigma_d=0.35$
kpc, which corresponds to the maximum spiral width arm observed for
HMSFRs~\citep{2014ApJ...783..130R} . The model evaluates $P(d)$ for
each source and compares it with a random value $k$ ($0<k<1$). If
$k> P(d)$, the source is rejected and the model takes another source
from the radial distribution to calculate $P(d)$ again and compare
it with a new $k$. However, if a source satisfies $k < P(d)$, then
the source is taken as a part of the model. The acceptance process
will continue until it reaches the total number of sources ($N$).
One example of a resulting spatial distribution can be seen in Figs.~\ref{samplot}
and~\ref{lv}.                                                           
 
\begin{figure} 
  \centering 
 \resizebox{\hsize}{!}{\includegraphics{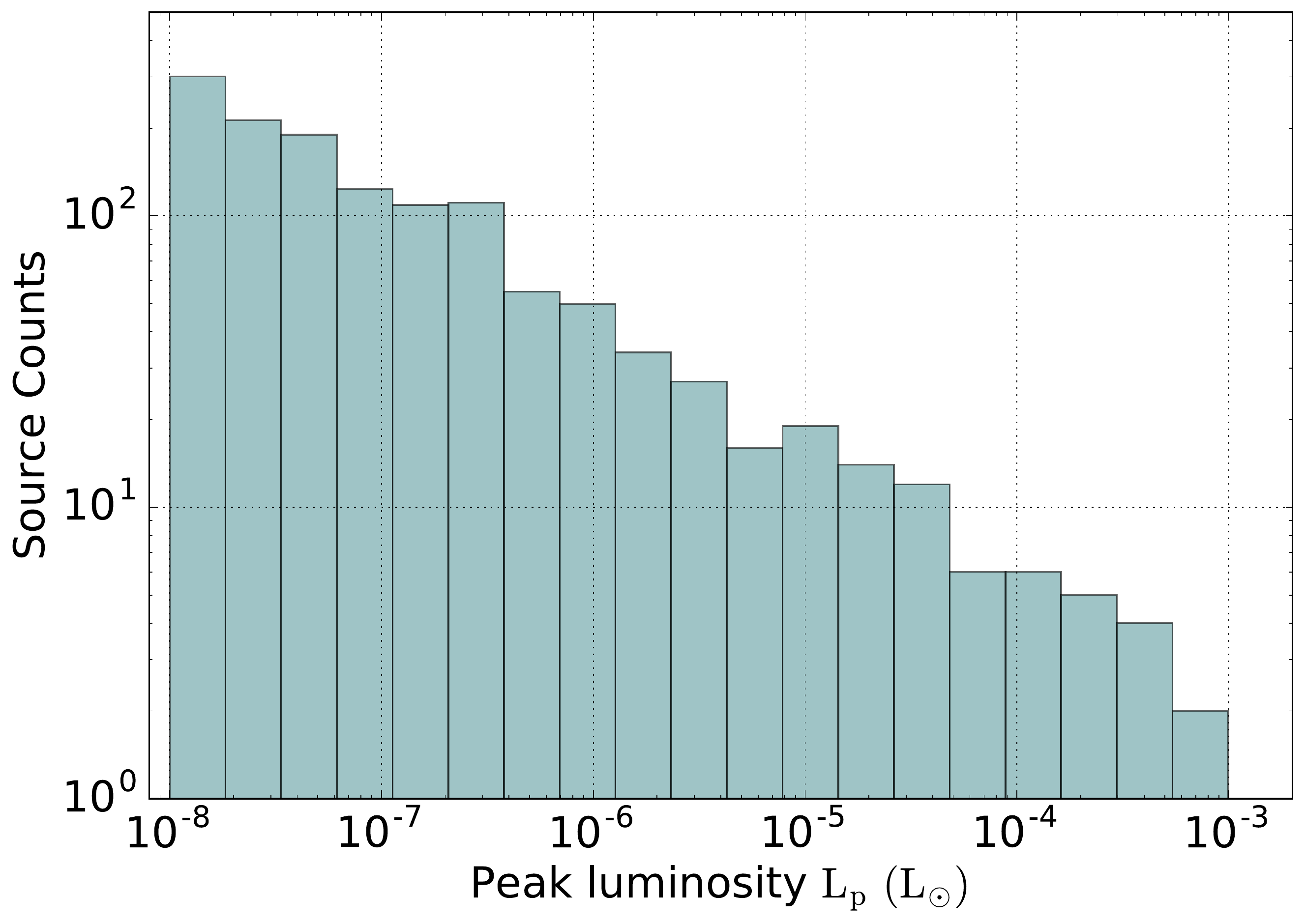}}                                                                    
 \resizebox{\hsize}{!}{\includegraphics{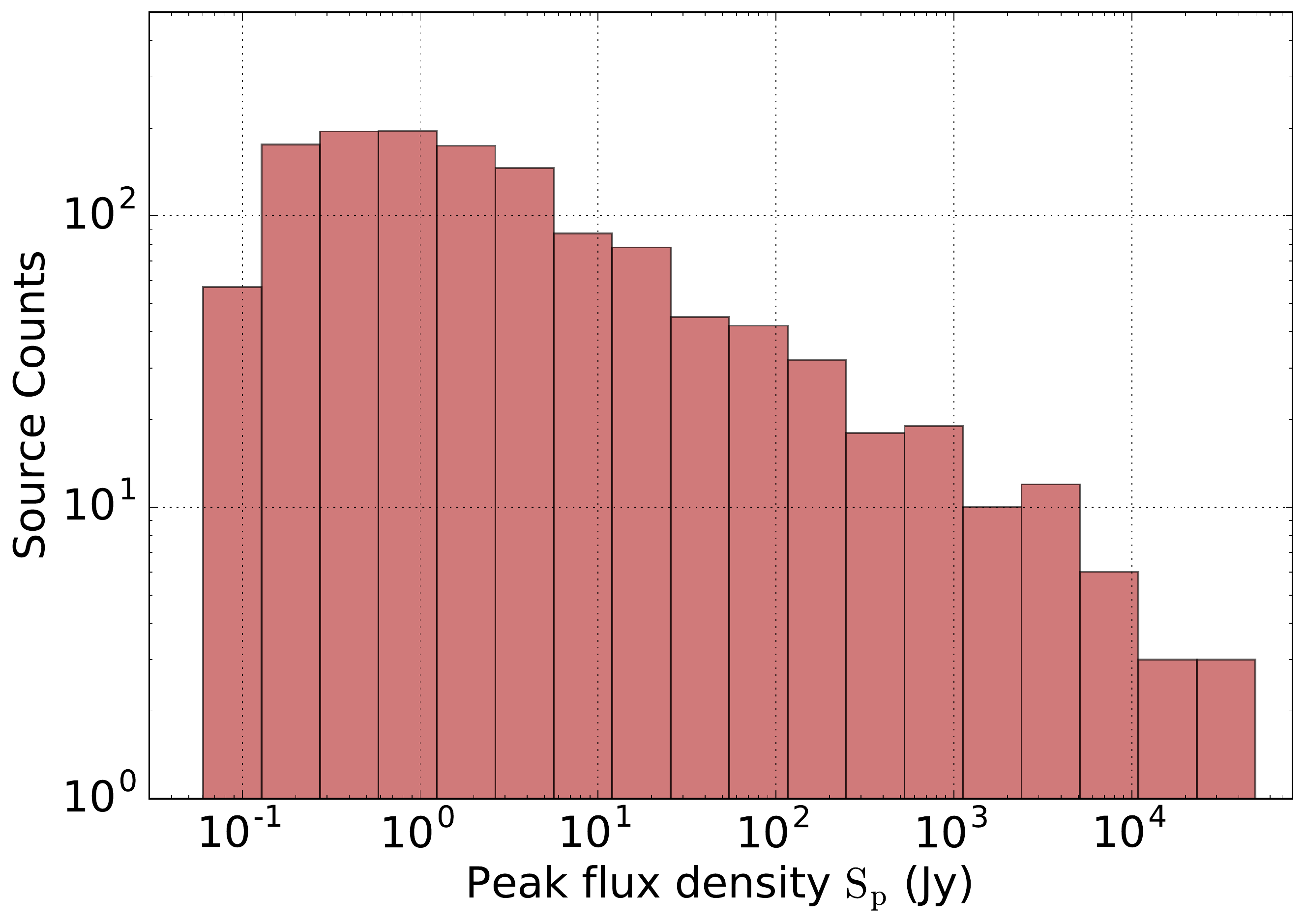}}                                                                   
  \caption{\label{luminosity_distr} \textbf{Top:} Peak luminosity
function adopted in the model using the fitted values for the total
number of 6.7 GHz methanol masers ($N=1300$) and the slope of the
luminosity function ($\alpha=-1.43$), see Sect.~\ref{ld} and~\ref{Luminosity_Results}.
\textbf{Bottom:} Peak flux density function obtained without sensitivity
limit.}                                                                 
\end{figure} 
 
\subsection{Velocity distribution} 
\label{vd} 
 
For the velocity distribution, we used a cylindrical coordinate system
($U,V,W$) in a rotational frame with an angular velocity of $\Theta
(R)$ in the direction of the Galaxy rotation, i.e., clockwise seen
from the NGP. In this system, $U$ is the radial component defined
positive towards the center of the Galaxy, $V$ is the tangential
velocity component defined positive in the direction of the Galactic
rotation and $W$ is the vertical velocity component defined positive
towards the NGP.                                                        
 
We drew Gaussian distributions for each velocity component
independently using the values, distributions and dispersions related
in Table~\ref{model_sum}. For the tangential velocity, we adopted
a Gaussian distribution with a mean value given by $\Theta (R)= \Theta_0
+ d\Theta/dR (R-R_0)$ and a dispersion of  $\sigma_{t}=$ 9 km $\rm{s^{-1}}$
(see Table~\ref{model_sum}). The values for $\Theta_0$, $d\Theta/dR$
and $R_0$ are provided in Table~\ref{initial}. The lower right panel of
Fig.~\ref{samplot} and Fig.~\ref{lv} show the distribution of
the Galactic tangential velocities $\Theta (R)$ and the maser velocities with respect
to LSR as a function of Galactocentric distance and Galactic
longitude respectively, assuming the values listed in Table~\ref{initial}.
The adopted dispersions for radial and vertical velocity components 
($\sigma_{r,v} = 5 \, \rm{km \ s^{-1}}$) are consistent with our estimates of
virial motions of individual massive stars, based on BeSSeL data,
whereas $\sigma_t$ was set larger to allow for the possible effects
of gravitational accelerations in the presence of material near spiral
arms.                                                                  
 
\subsection{Methanol masers represented in the model} 
 
The BeSSeL survey determined proper motions and parallaxes of water
masers (at 22 GHz) and methanol masers (at 6.7 and 12 GHz) and fit
them to an axially symmetric Galactic model to estimate the Galactic
parameters. Compared with the BeSSeL survey, we have made a simplification
by assuming that all sources are selected from 6.7 GHz methanol masers
surveys, but observed with VLBI at 12 GHz.                              
 
\subsection{Luminosity distribution} 
\label{ld} 
 
Notably, for our model it is important to estimate astrometric observational
errors based on maser detectability, which are directly related to
the peak flux density ($S_p$) of each maser, i.e., the flux density
emitted in a specific line integrated over a single channel width.
The peak flux density function can be estimated if the
peak luminosity function and the spatial distribution are known, assuming isotropic
emission. Although the individual maser spots may not radiate
isotropically, we assume that this holds over the sample of randomly
oriented masers.                                                        
 
\cite{2007A&A...463.1009P} have suggested that the 6.7 GHz methanol
maser luminosity distribution takes the form of a single power law
with sharp cutoffs of $10^{-8} \, \rm{L_{\sun}}$ and $10^{-3} \,
\rm{L_{\sun}}$ and a slope ($\alpha$) between $-1.5$ and $-2$. We
assume the same dependence for the peak luminosity function (see
Fig.~\ref{luminosity_distr}), but we refine it by varying the parameters
to match the results of the MMB survey. The results of this procedure are presented in Sect.~\ref{Luminosity_Results}.
 
\subsection{Error allocation} 
\label{allocation} 
 
In order to be able to use simulated data in tests to estimate
the Galactic parameters, it is necessary to assign observational
error distributions. For our model, the errors in the parallax and
proper motions were estimated following a calculation for relative
motions of maser spots and statistical parallaxes, i.e., $\sigma_{
\pi} \propto \Theta_{res}/ (S/N)$ and $\sigma_{ \mu_{\alpha,\delta}}
= \sigma_{ \pi}/(1\rm{yr})$, where $\Theta_{res}$ is the VLBA resolution
for 12 GHz methanol masers. The signal-to-noise ratio ($S/N$) depends
on the peak flux density value ($S_p$) and given that most of the
current data of the BeSSeL survey are based on VLBA observations,
we adopted a channel width of 50 kHz ($1.24 \, \rm{km \, s^{-1}}$)
at 12 GHz and an integration time of 2 hr. This was used to estimate
the $S/N$ and thus the errors in parallax and proper motions. \cite{2014ApJ...783..130R}
estimated an additional error term for $\sigma_{V_{los}}$ ($5 \,
\rm{km \, s^{-1}}$), which is associated with the uncertainty on
transferring the maser motions to the central star. This error 
dominates the BeSSeL observations of $V_{los}$, and this uncertainty
is reflected in the value of $\sigma_{V_{los}}$.                        
 
\begin{figure} 
 \resizebox{\hsize}{!}{\includegraphics{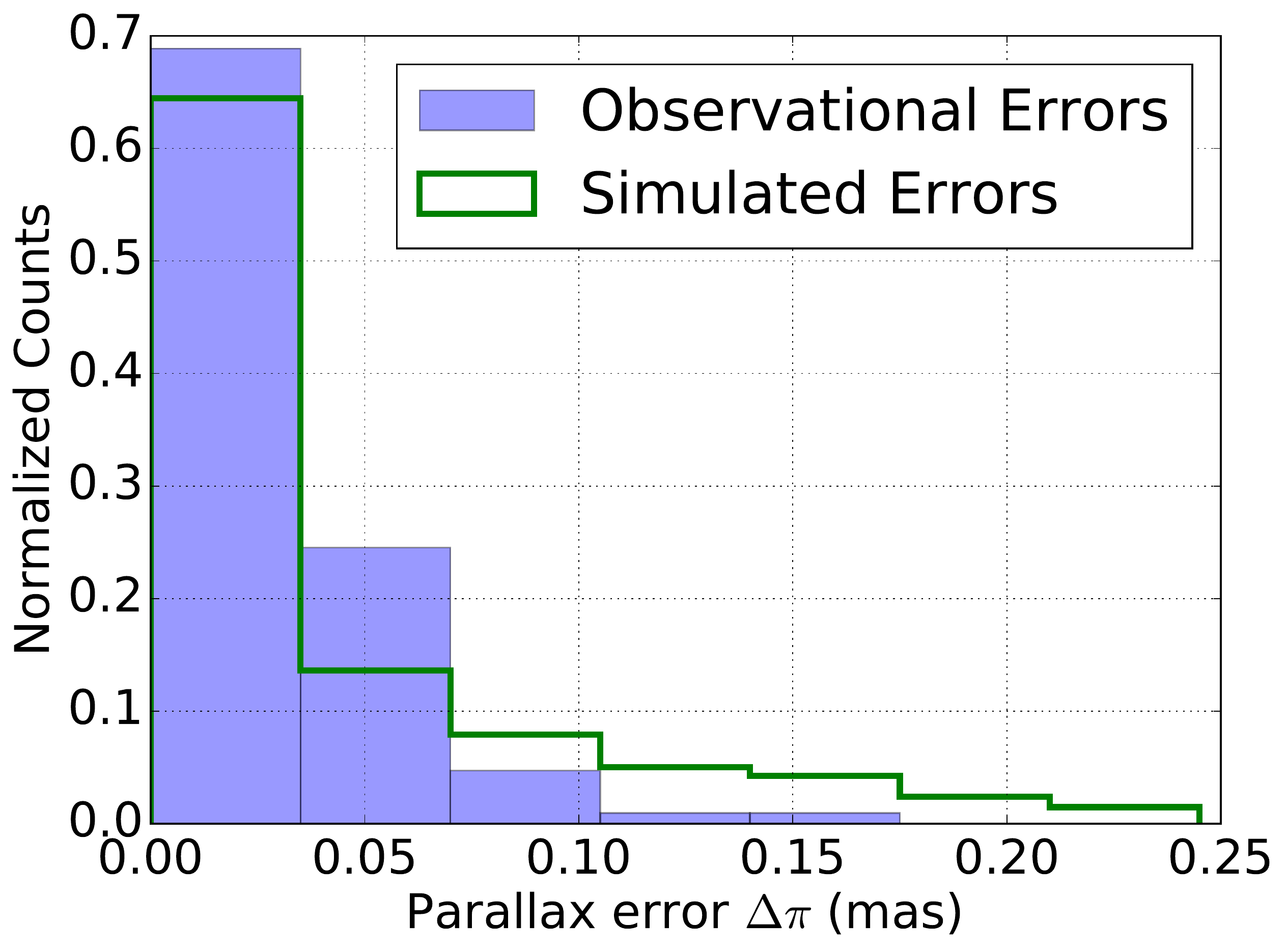}} 
  \caption{\label{err}Comparison between the error distribution for
observational and simulated parallax measurements. Observational
errors are based on 103 astrometric sources published in~\cite{2014ApJ...783..130R}
as part of the BeSSeL survey.}                                          
\end{figure}

Parallax estimates in~\cite{2014ApJ...783..130R} are often dominated
by residual, whereas troposphere-related errors dominate in the astrometry, and so we adopted
a simple prescription for parallax uncertainty (as shown above),
which does not directly include systematic effects. However, when a large number of simulated sources are used, many weak masers are
included that would be $S/N$ limited. Figure~\ref{err} shows a comparison
between the two error distributions for observational and simulated
parallax measurements in which our $S/N$ error estimate 
yields a similar distribution to the uncertainties used in~\cite{2014ApJ...783..130R}.
 
The fitting procedure described by~\cite{2014ApJ...783..130R} used
to determine the Galactic parameters (combining BeSSeL and VERA data)
requires high accuracy VLBI data as input. This data consists of a 3D position vector ($\alpha$, $\delta$, $\pi$), a 3D velocity vector ($\mu_{\alpha}$, $\mu_{\delta}$,$V_{los}$),
and the errors $\sigma_{\pi}$, $\sigma_{ \mu_{\alpha}}$, $\sigma_{\mu_{\delta}}$
and $\sigma_{V_{los}}$. Although the model gives exact values for
position and velocities of each maser source seen from the Earth,
we are interested in realistic values as input for the fitting procedure.
Therefore, we add a noise component to each observable
quantity ($\pi$, $\mu_{\alpha}$, $\mu_{\delta}$, $V_{los}$), using 
random values following Gaussian distributions with standard deviations
equal to the estimated errors previously calculated. By changing
the error distribution, we can control the quality of the data entered
in the fitting procedure.                                               
 
\begin{table} 
\centering 
\resizebox{\hsize}{!}{ 
\begin{tabular}{c c c c c} 
\hline\hline 
 & \multicolumn{2}{c}{MMB}  &  \multicolumn{2}{c}{Arecibo}  \\ 
&Observation&Simulation&Observation&Simulation\\ 
\hline 
Sensitivity (3$\sigma$)& \multicolumn{2}{c}{$\le 0.71$ Jy} &\multicolumn{2}{c}{$\le
0.27$ Jy} \\                                                            
\hline 
Sky & \multicolumn{2}{c}{$-174^{\circ}\le l \le 60^{\circ}$} & \multicolumn{2}{c}{$35.2^{\circ}
\le l \le 53.7^{\circ}$}\\                                              
coverage & \multicolumn{2}{c}{$-2^{\circ} \le b \le 2^{\circ}$} &
\multicolumn{2}{c}{$-0.4^{\circ} \le b \le 0.4^{\circ}$}\\              
\hline 
Sources&908&800$\pm$20&76&95$\pm$10\\ 
\hline 
$\beta$ &-0.60$\pm$0.1&-0.44$\pm$0.1&-0.36$\pm$0.1&-0.38$\pm$0.1\\ 
\hline 
\end{tabular}}
\caption{\label{limitations} Limits in sensitivity and source location,
numbers of masers detected and the slope of the flux density functions
($\beta$) for the 6.7 GHz methanol masers surveys: MMB and Arecibo.
Limits of both surveys, numbers of masers ($N$) and slope of the
luminosity function ($\alpha$) fitted in Sect.~\ref{Luminosity_Results},
were applied to our Galactic model; the results are displayed in
the columns labeled "Simulation". The simulated errors correspond
to the standard deviation after running 100 simulated galaxies.}        
\end{table} 
 
\begin{figure} 
 \resizebox{\hsize}{!}{\includegraphics{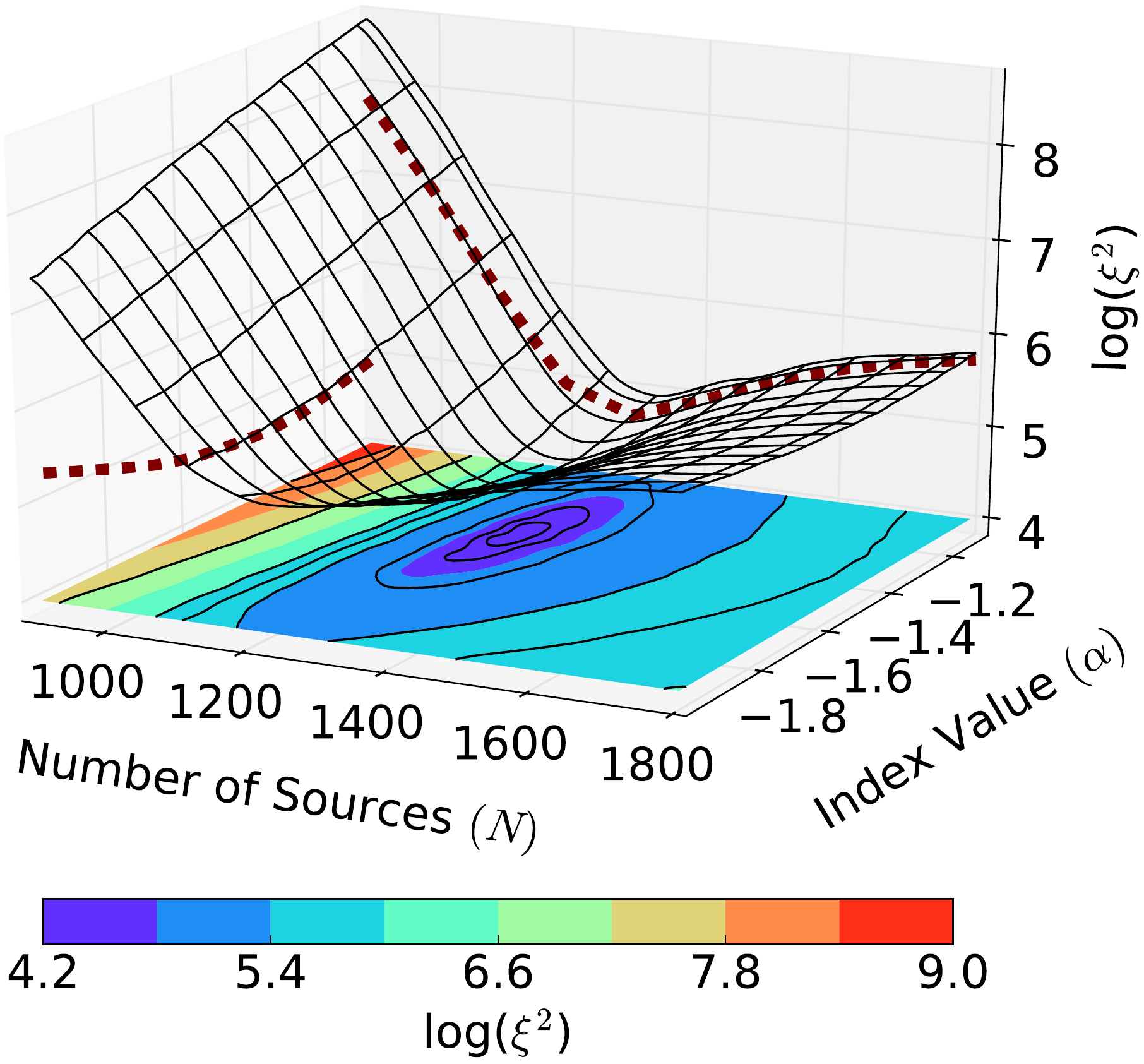}} 
  \caption{\label{minimization}Grid of initial parameters displaying
the $\xi^2$ calculation for each $N$, $\alpha$ pair. The dark
blue region represents the best values of $N$ and $\alpha$ that most closely match the MMB results. The projected gray dashed lines show the
profiles of the surface close to the minimum values of $\xi^2$.}        
\end{figure} 
 
\subsection{Fitting procedure} 
\label{fitting} 
 
The fitting procedure used was adopted from the BeSSeL survey~\cite[see][]{2009ApJ...700..137R,2014ApJ...783..130R}.
The input data for the fitting procedure are 3D position and 3D velocity
information of the masers, conservative priors for the solar motion,
the average source peculiar motion, and the Galactic scale and rotation.
Convergence on the best Galactic parameters to match the spatial-kinematic
model was made using a Bayesian fitting approach, where the velocities
were used as known data to be fitted, and the sky positions and distances
were used as coordinates. The posterior probability density function
(PDF) of the Galactic parameters were estimated with Markov chain
Monte Carlo (MCMC) trials that were accepted or rejected by a Metropolis-Hastings
algorithm~\citep[see][for a detailed explanation]{2009ApJ...700..137R,2014ApJ...783..130R}.
Finally, the procedure returns the best Galactic parameter values
that match the simulated data to the spatial-kinematic model. The
fitting procedure was improved compared to that used in~\cite{2009ApJ...700..137R,2014ApJ...783..130R}:
first, the fitting procedure now corrects for bias when inverting
parallax to estimate distance, which becomes significant when fractional
parallax uncertainties exceet $\approx15$\%~\citep[note this is not a trivial inference problem,
see e.g.][]{2015PASP..127..994B}; second, the fitting procedure
was improved by adding a term to the motion uncertainties, which comes
from parallax uncertainty. After these two modifications, the fitting
procedure yielded unbiased Galactic parameter values, even when weak and/or 
very distant masers with large fractional parallax uncertainties were simulated.
 
\section{Results} 
\label{results} 
 
\begin{figure} 
 \resizebox{\hsize}{!}{\includegraphics{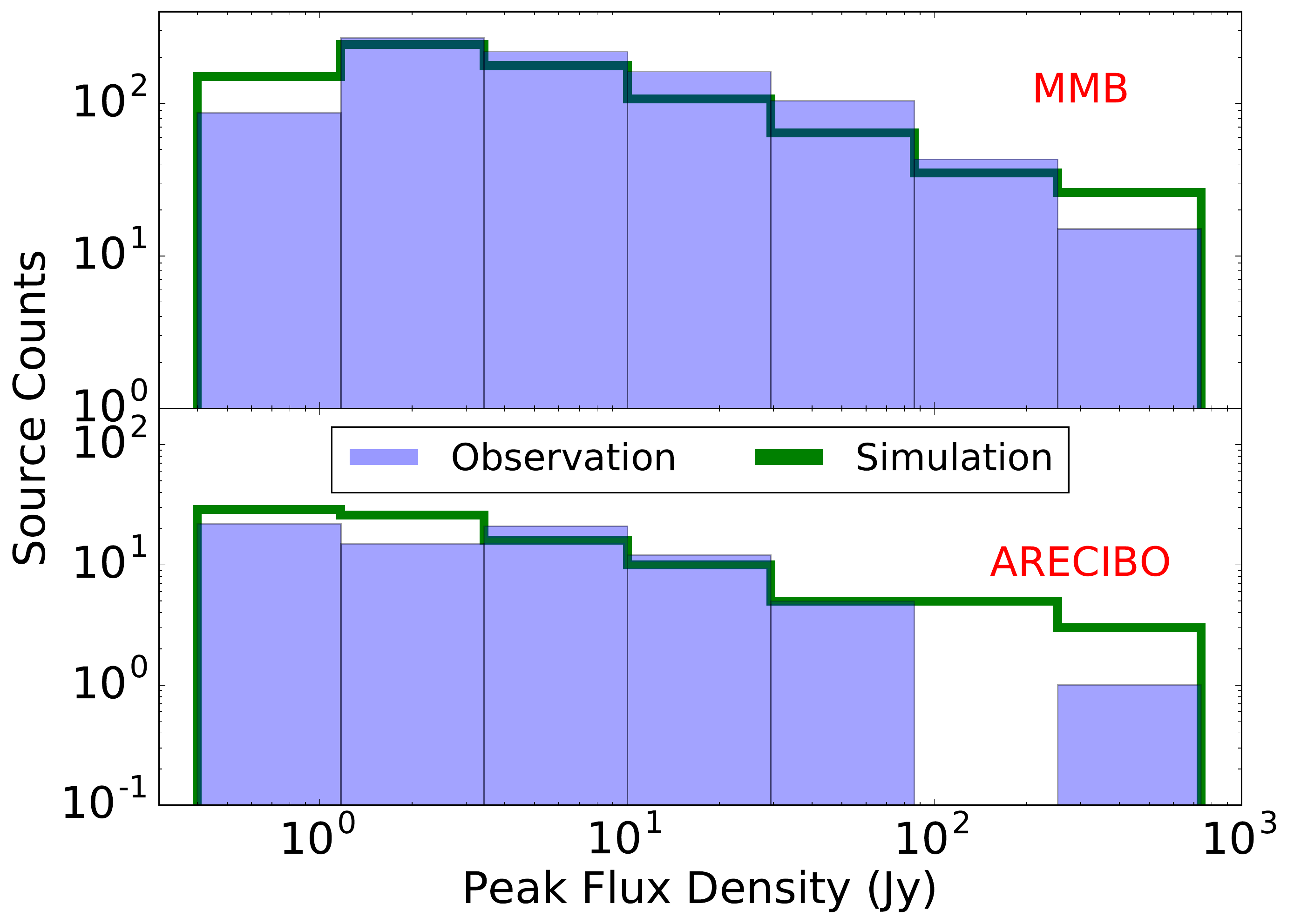}} 
  \caption{\label{fluxdensityMMBAr} In blue: flux density function
obtained for the MMB (top) and the Arecibo survey (bottom). In green:
simulated flux density function obtained in the model (using $N=
1300$ and $\alpha= -1.43$) after the MMB and Arecibo limits were
applied (Table~\ref{limitations}).}                                     
\end{figure} 
 
\begin{figure*}%
  \centering 
  \resizebox{\hsize}{!}{\includegraphics{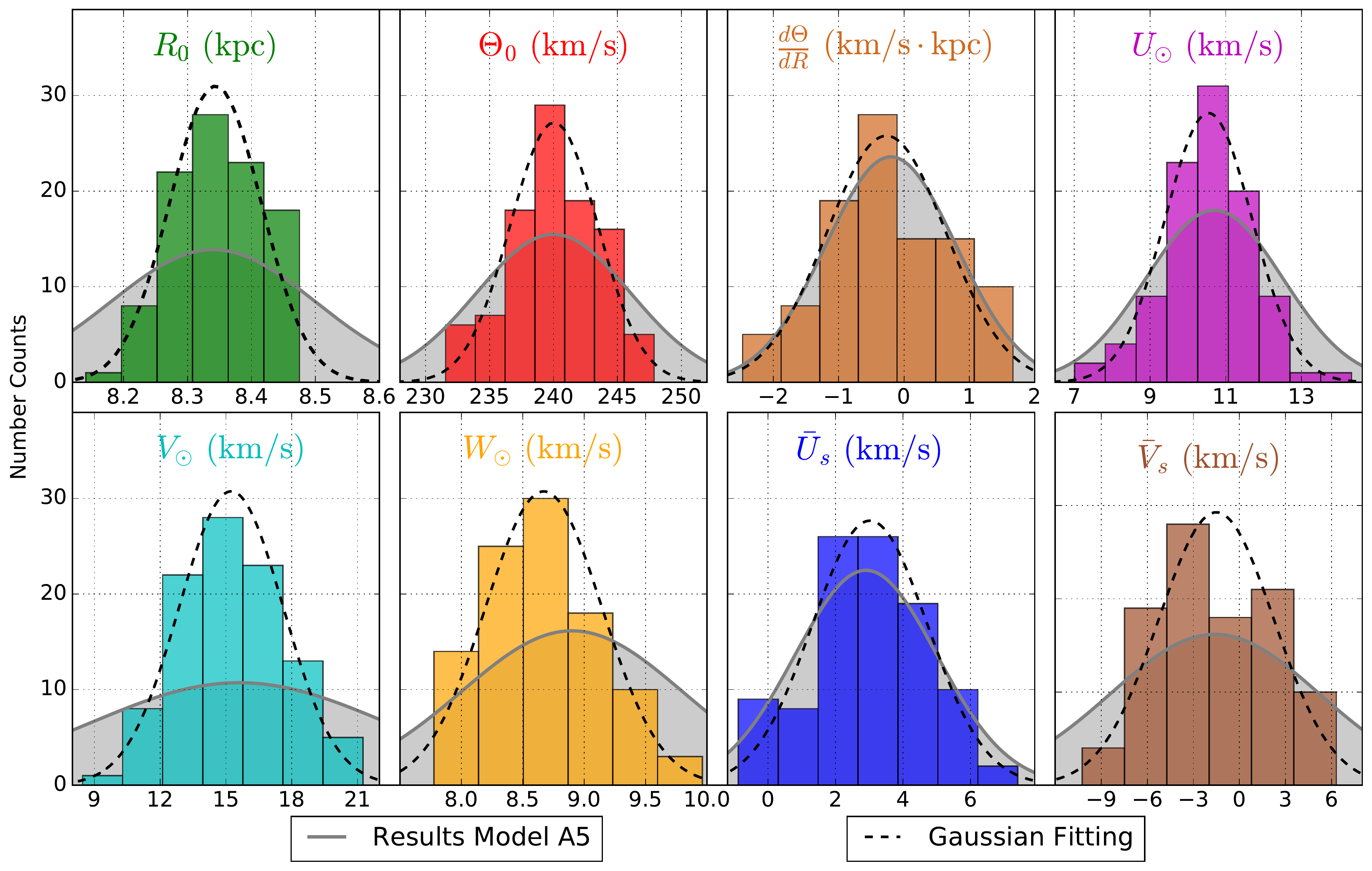}} 
  \caption{\label{eight} Galactic parameters distributions found
for 100 simulated galaxies mimicking the BeSSeL data sample selection
(Sect.~\ref{GP_Results}). The values listed in Table~\ref{Model_BeSSeL}
correspond to the fitting made to the histograms and shown as black
dashed lines. Bayesian fitting results for the A5 model reported in~\cite{2014ApJ...783..130R}
are shown as gray regions.}                                             
\end{figure*}%
 
\begin{figure*}%
  \centering 
  \resizebox{\hsize}{!}{\includegraphics{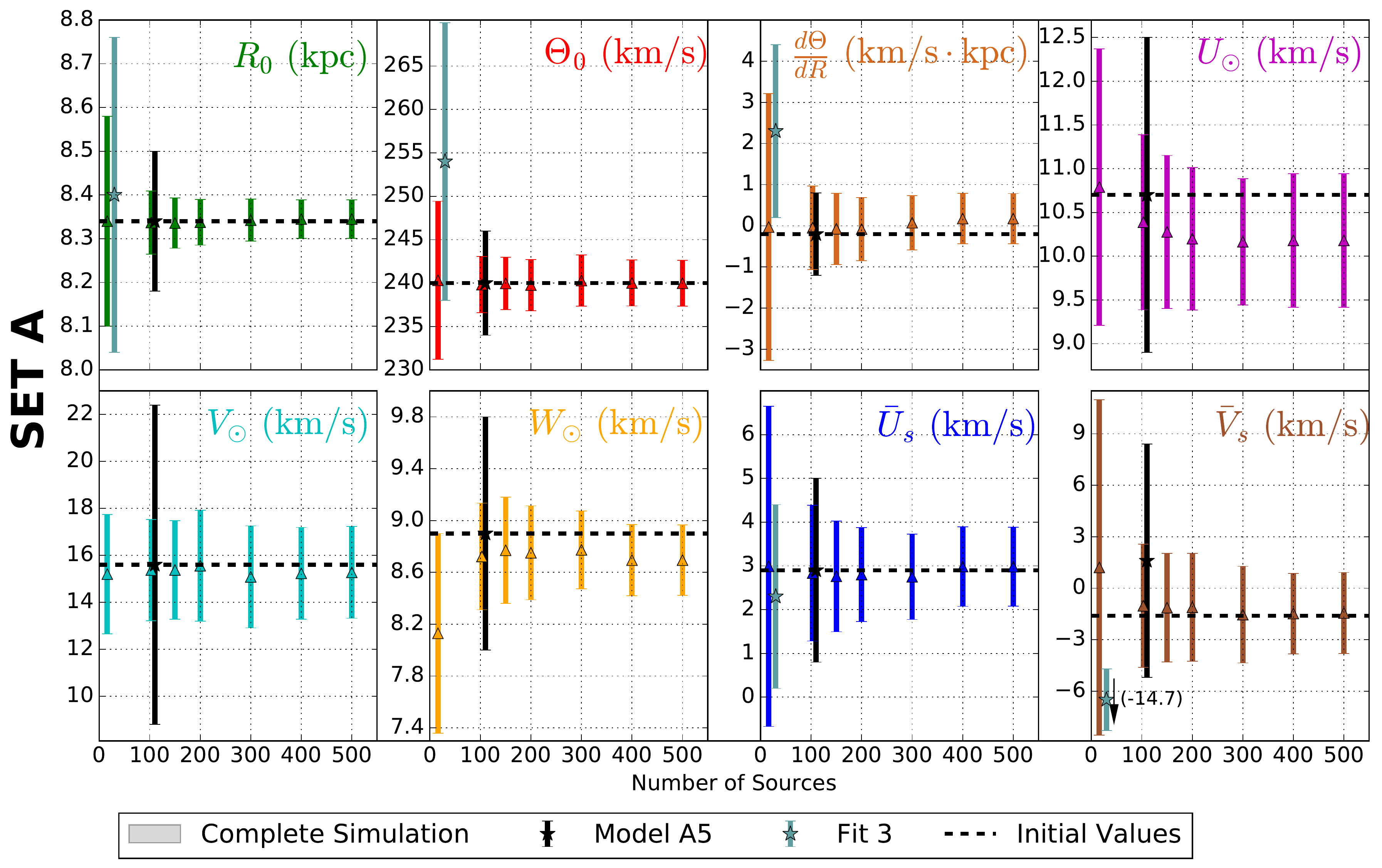}} 
  \resizebox{\hsize}{!}{\includegraphics{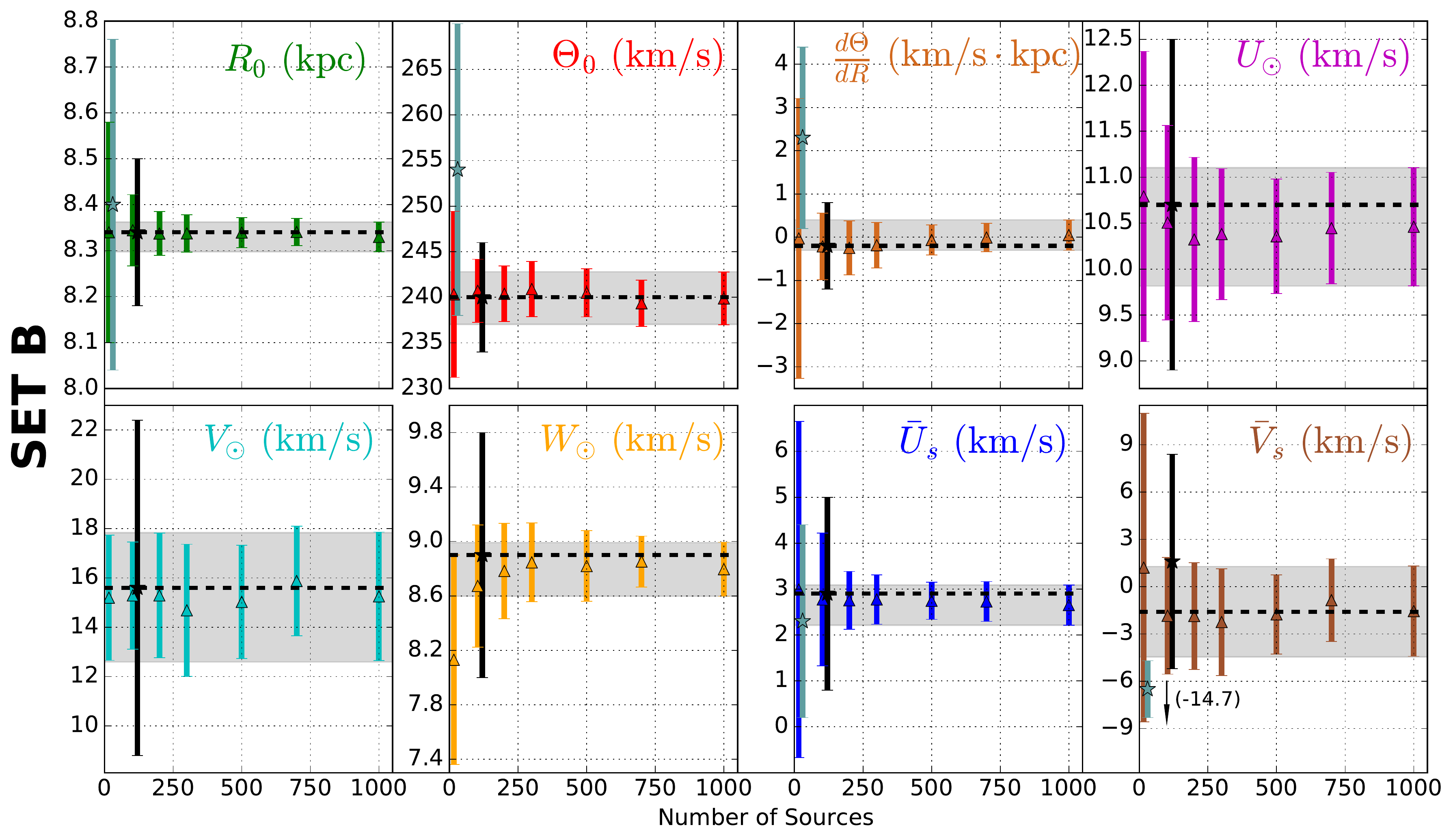}} 
\caption{\label{plot_moresources_evolution} Galactic parameter values
obtained for samples in sets A and B. In each sample, sources are
added in the northern hemisphere simulating the future BeSSeL results
(set A) and without location limit simulating samples when southern
arrays can contribute with data (set B). First and current BeSSeL
results published in~\cite{2009ApJ...700..137R,2014ApJ...783..130R},
respectively labeled "Fit 3" and "Model A5", are shown as stars for
comparison. The initial values adopted in the model are represented
as dashed lines. Gray regions correspond to values and uncertainties
obtained for the complete sample ($N=1300$).}                           
\end{figure*}%
 
A comparison of the systematic 6.7 GHz methanol maser observational
surveys and the simulated model peak flux density is shown in Sect.~\ref{Luminosity_Results}.
In Sect.~\ref{GP_Results}, different sample selections are used to
compare the Galactic parameters obtained with respect to the initial
values used (Table~\ref{initial}). Finally, in Sect.~\ref{correlations},
the Pearson correlation coefficients are calculated to quantify correlations
among the Galactic parameters.                                          
 
\subsection{Luminosity function for 6.7 GHz methanol masers} 
\label{Luminosity_Results} 
 
We compared the flux density distribution functions of the MMB survey
and the Arecibo survey with the current model to fit two parameters:
the total number of sources ($N$) and the slope of the peak luminosity
function ($\alpha$). The MMB survey is the most sensitive
unbiased survey yet undertaken for 6.7 GHz methanol masers. The Parkes
Observatory was upgraded with a seven-beam receiver to carry
out a full systematic survey of the Galactic Plane~\citep[][and the references
within]{2012MNRAS.420.3108G}.  The Arecibo Survey was a deep
a 6.7 GHz methanol maser survey over a limited portion of the Galactic
plane~\citep{2007ApJ...656..255P}.
 
Table~\ref{limitations} summarizes the survey limits in sensitivity
and sky coverage for the MMB and Arecibo surveys. The last two rows
list the number of sources detected and the slope of the flux density function
($\beta$) for each survey. By using these data, we were able to make
a direct comparison between the simulated and
observed flux density functions for each survey (green and blue
histograms in Fig.~\ref{fluxdensityMMBAr}). For our comparison, we
excluded MMB sources that reside inside a Galactocentric radius of 3 kpc as this region is
not part of the model.                                                  
 
In order to fit $N$ and $\alpha$ to the results of the 
surveys, a grid of initial parameters (Fig.~\ref{minimization}) was
sampled using similar ranges to those proposed by~\cite{2007A&A...463.1009P} for $N=[900,1800]$ and~\cite{2005MNRAS.360..153V} for $\alpha=[-1.1,2.0]$. The grid was constructed
such that each point represents a pair of initial parameters ($N$,
$\alpha$) and for each pair, a set of simulated galaxies was generated
following the initial conditions described in Sect.~\ref{model}.
Next, the surveys limits (Table~\ref{limitations}) were applied,
and we compared the flux density function obtained for each $N$, $\alpha$ pair with the flux density function of the MMB survey (blue histogram
in the top of Fig.~\ref{fluxdensityMMBAr}). Through a minimization
procedure, we found values of $N$ and $\alpha$ that best match the MMB results.
This procedure was implemented only for the MMB data since it represents
a larger and more complete sample than the Arecibo survey. The minimization
procedure compares the MMB observed (blue) and the simulated (green)
flux density functions (see Fig.~\ref{fluxdensityMMBAr}) and minimizes
a quantity called $\xi^2$, where                                 
 
\begin{equation} 
\xi^2 = \sum_{bins} \frac{(y-y_{obs})^2}{y_{obs}}, 
\end{equation} 
 
\noindent and $y$ represents the number of sources per luminosity
bin. Given that our Galactic model generates galaxies based on a
stochastic method, the position, velocity and luminosity values for
each maser vary each time the model is executed (even using
the same pair of $N$ and $\alpha$). By generating sets of ten independent
galaxy simulations per $N$, $\alpha$ pair, we found that the
fluctuations in the simulations were smaller than the uncertainties
in the binned data, and hence this procedure was applied.               
 
Figure~\ref{minimization} shows the values obtained for $\xi^2$ per $N$, $\alpha$ pair as a 3D surface. The dark blue region in
the projected contour plot represents the best set of parameters
that mimic the MMB survey results.
We found that the surface near the minimum
can be approximated by a Gaussian in two dimensions (see projections
in Figure~\ref{minimization}). Using the maximum likelihood estimation,
which is well defined for multivariate Gaussian distributions, we
estimated the mean and its respective uncertainty. The best parameters
were found to be $N=1300 \pm 60 $ sources and $\alpha = -1.43 \pm
0.18$. Finally, Fig.~\ref{fluxdensityMMBAr} shows the flux density
function for the MMB (top), and Arecibo survey (bottom) in blue,
and their respective simulated flux density function are shown in
green for the best parameters of $N$ and $\alpha$ found.
Additionally, the number of sources detected and the slope of the
flux density function ($\beta$) for the simulated surveys are listed
in Table~\ref{limitations}.                                             
 
\subsection{Galactic parameters and selection of sample} 
\label{GP_Results} 
 
\begin{table} 
\centering 
\resizebox{\hsize}{!}{ 
\begin{tabular}{c c c} 
\hline \hline 
Galactic & Simulated & A5 \\ 
Parameter & BeSSeL Sample & Model \\ 
\hline 
$R_0 \, \rm{(kpc)}$  &$ 8.34 \pm 0.07$ & $8.34 \pm 0.16$ \\ 
$\Theta_0  \, \rm{(km \, s^{-1})} $&$ 240.0 \pm 3.4$ & $240.0 \pm
8.0 $ \\                                                                
d$\Theta$/d$R  \, \rm{(km \, s^{-1} \, kpc^{-1})}$&$ -0.3 \pm 0.9$
& $-0.2 \pm 0.4$\\                                                      
$U_{\sun} \, \rm{(km \, s^{-1})}$&$ 10.5 \pm 1.2$ & $10.7  \pm 1.8 $ \\ 
$V_{\sun}  \, \rm{(km \, s^{-1})} $&$ 15.2 \pm 2.4$ & $15.6  \pm
6.8 $ \\                                                                
$W_{\sun}  \, \rm{(km \, s^{-1})} $&$ 8.7 \pm 0.5$ & $8.9  \pm 2.1 $ \\ 
$U_s  \, \rm{(km \, s^{-1})} $&$ 3.0 \pm 1.7$ & $2.9  \pm 2.1 $ \\ 
$V_s  \, \rm{(km \, s^{-1})} $&$ -1.5 \pm 3.8$ & $-1.5  \pm 6.8 $ \\ 
\hline 
\end{tabular}} 
\caption{\label{Model_BeSSeL} Galactic parameter results for 100
simulated galaxies mimicking the BeSSeL data sample. Additionally,
the Bayesian fitting results for the A5 model reported in~\cite{2014ApJ...783..130R},
which were also the initial values adopted in the model (Table~\ref{initial}),
are shown for comparison.}                                              
\end{table} 
 
\begin{figure} 
\resizebox{\hsize}{!}{\includegraphics{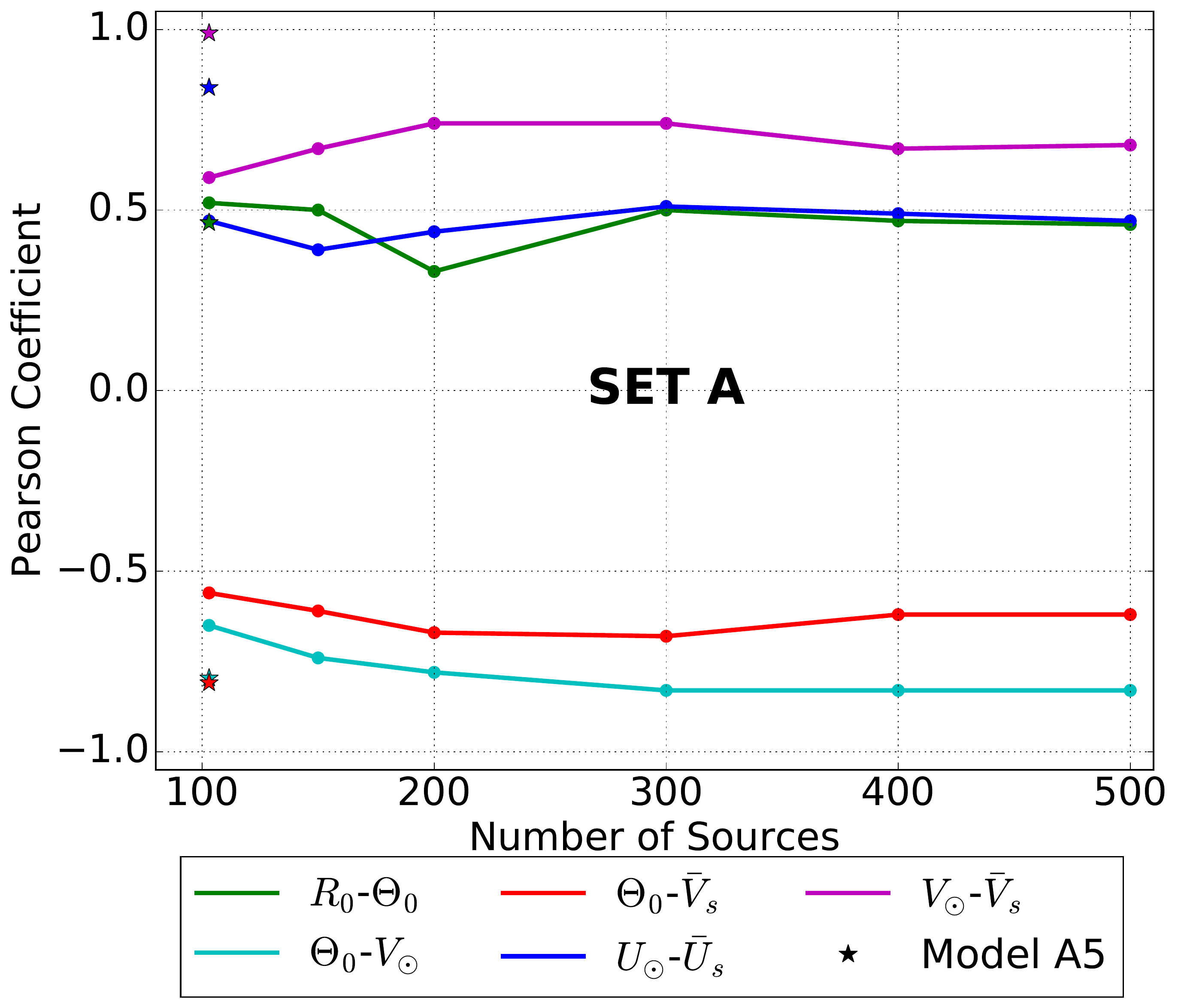}} 
\resizebox{\hsize}{!}{\includegraphics{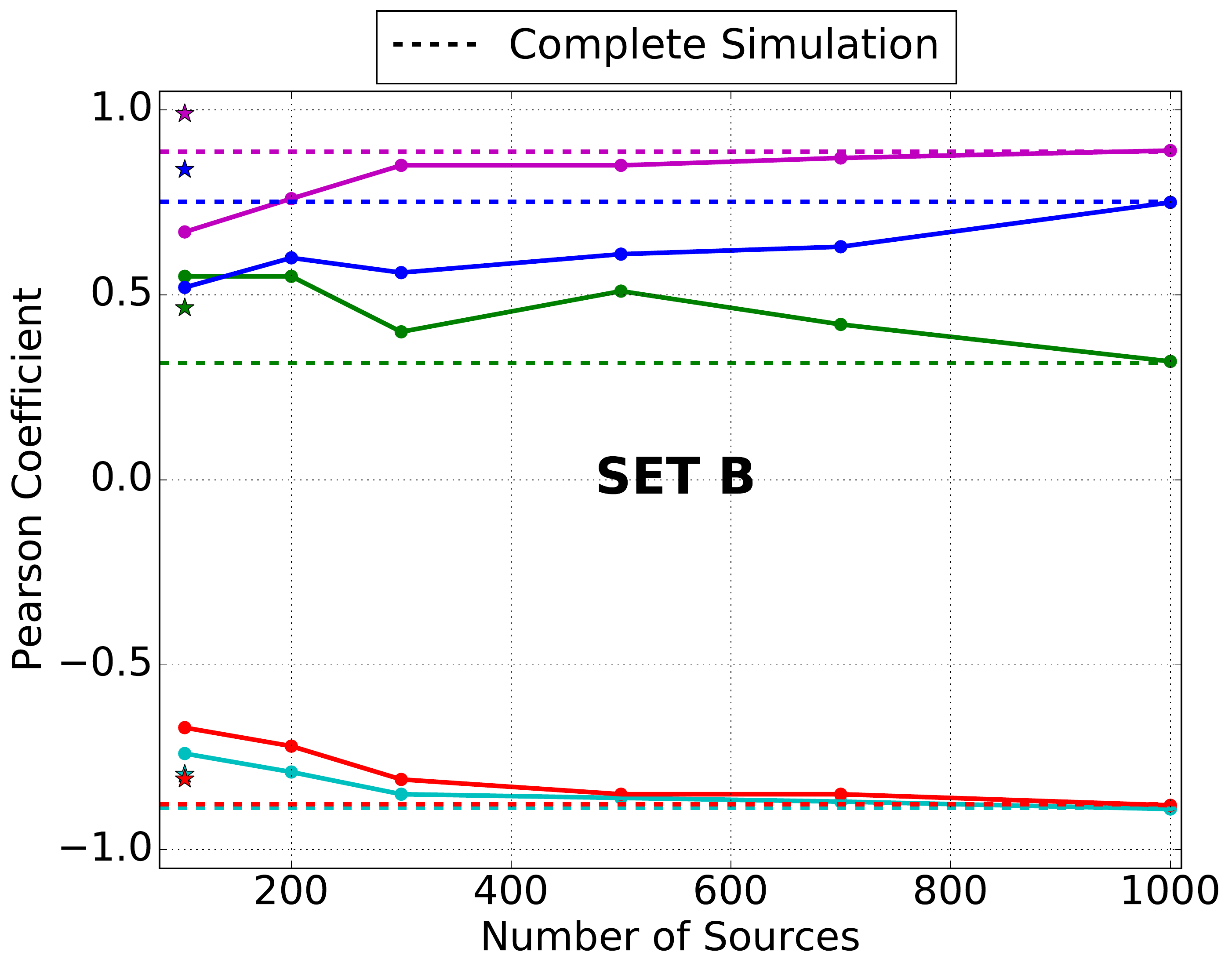}} 
  \caption{\label{pearson_evol} Pearson product-moment correlation
coefficients calculated for initially highly correlated values (see
Table~\ref{pearsons}), when more sources are added in sets A (top
panel) and B (bottom panel). Pearson coefficients reported by~\cite{2014ApJ...783..130R}
are shown as stars and those for the complete sample ($N=1300$) are
represented by dashed lines.}                                           
\end{figure} 
 
The model can reproduce the methanol maser distribution for the entire
Galaxy including observational errors. In order to evaluate
the possible biases introduced by the observed BeSSeL sample (equivalent
to the 103 brightest sources in the declination region, $-30^{\circ}
\le \delta \le70^{\circ}$, which is equivalent to $-2^{\circ} \le
l \le242^{\circ}$), 100 galaxies were simulated to mimic the BeSSeL
sample. Then, they were fitted to test whether the adopted Galactic
parameters were returned. Figure~\ref{eight} shows the distribution
obtained on each Galactic parameter for the simulated BeSSeL sample
compared with the values reported in~\cite{2014ApJ...783..130R}.
The histograms were fitted to Gaussian distributions, and the results
are shown in Table~\ref{Model_BeSSeL}. Clearly, in all cases the
distributions of fitted values are centered on the adopted value,
and in most cases the widths of the distributions are smaller than those
reported in~\cite{2014ApJ...783..130R}.                                 
 
\begin{table*}%
\centering 
\resizebox{\hsize}{!}{ 
\begin{tabular}{c c c c c c c c c} 
\hline 
\hline 
 & $R_0$ &  $\Theta_0$ & d$\Theta$/d$R$ & $U_{\odot}$ & $V_{\odot}$
& $W_{\odot}$ & $\bar{U}_s$ & $\bar{V}_s$ \\                            
\hline 
$R_0$&1.00 (1.00)&&&&&&&\\ 
$\Theta_0$&0.48 (0.47)&1.00 (1.00)&&&&&& \\ 
d$\Theta$/d$R$&0.27 (0.10)&0.21 (0.14)&1.00 (1.00)&&&&& \\ 
$U_{\sun}$&0.32 (0.45)&0.01 (0.24)& 0.08 (-0.12)&1.00 (1.00)&&&& \\ 
$V_{\sun}$&0.02 (0.02)&-0.73 (-0.80)&0.05 (-0.01)&0.05 (-0.01)&1.00
(1.00)&&& \\                                                            
$W_{\sun}$&-0.01 (0.00)&0.02 (-0.01)&-0.01 (0.03)&-0.06 (-0.02)&0.01
(0.01)&1.00 (1.00)&& \\                                                 
$\bar{U}_s$&0.51 (0.52)&-0.09 (0.17)&0.01 (-0.09)&0.51 (0.84)&0.05
(-0.01)&-0.04 (0.00)&1.00 (1.00)& \\                                    
$\bar{V}_s$&0.09 (0.00)&0.62 (0.81)&-0.47 (0.02)&-0.16 (-0.03)&-0.68
(-0.99)&0.01 (-0.01)&0.13 (0.03)& 1.00 (1.00)\\                         
\hline 
\end{tabular}}%
\caption{\label{pearsons} Pearson product-moment correlation coefficients
calculated for 100 galaxies simulated to mimic the BeSSeL data
sample selection. The respective Pearson coefficient
reported in~\cite{2014ApJ...783..130R} for the observed sample are
listed in parentheses.}                                                                
\end{table*}%
 
In addition to the 100 simulated galaxies that mimic the BeSSeL sample,
we also simulated the first BeSSeL data sample, where only 16 HMSFRs
over the northern hemisphere were used to estimate the same Galactic
parameters but not the solar motion~\citep{2009ApJ...700..137R}.
Moreover, we also started adding sources to form two additional sets
of simulated data. Set A was made to study the impact of future viable
observations with the VLBA, EVN, and VERA to obtain up to 500 sources
in the northern hemisphere. Again, we selected the brightest sources
first to fall in the same declination range that BeSSeL is targeting
for this. We generated samples from 16 up to 500 sources, which were
drawn from the total number sources ($N=1300$) that may lie in the
declination range proposed. Set B represents the conditions for a
more complete effort when VLBI arrays in the Southern hemisphere
can contribute to the astrometric sample. As was done in set A, we selected
the brightest sources but now without declination limitation, generating
samples from 16 up to the complete sample ($N=1300$). Each additional
sample in both sets was simulated for 100 galaxies. We note that in
all cases the errors continued to be based on the VLBA observations
characteristics.                                                        
 
Figure~\ref{plot_moresources_evolution} shows how the Galactic parameter
values change as more sources are added to the sample selection for
sets A and B. The dashed lines represent the initial values adopted
in the model, and the error bars represent the standard deviation
found for each parameter. The first and current BeSSeL results are
also shown as stars and labeled following the same convention used
in~\cite{2009ApJ...700..137R} and~\cite{2014ApJ...783..130R}, i.e.,
Fit 3 and Model A5 respectively.                                        
 
Our objective was to investigate the accuracy with which the Galactic 
parameters can be recovered in the presence of measurement errors. It was 
therefore important that we verify the robustness of the fitting algorithm and its 
dependence on the choice of initial parameters. To make sure the fitting 
procedure recovers the Galactic parameters in an unbiased way over a large 
range, we ran the algorithm over a number of values in the multi-dimensional 
parameter space that defines our Galactic models. We varied the most relevant
parameters over a broad range ($\pm \Delta$ and $\pm 3 \Delta$ for the 
obtained simulated BeSSeL values related in Table~\ref{Model_BeSSeL}) 
and calculated a normalized difference between the input parameters and
the returned fits. We found that indeed the fitting procedure can properly 
recover the starting values.   
 
\subsection{Parameter correlations} 
\label{correlations} 
 
Using the Galactic parameter values obtained for 100 simulated galaxies
mimicking the BeSSeL data sample selection, we calculated the Pearson
product-moment correlation coefficients between all the parameters
from the output distributions. The coefficients found are shown in
Table~\ref{pearsons}; for comparison, the Pearson coefficient
estimates reported in~\cite{2014ApJ...783..130R} from the fitting
procedure are also listed. Pearson coefficients in~\cite{2014ApJ...783..130R}
were calculated by MCMC trials, but in our case we have a large number
of samples, which provides an independent way to estimate the correlations.
Our findings seems to be consistent with the coefficients published
in~\cite{2014ApJ...783..130R}.

We also estimated the Pearson coefficients variation as more sources
are added to the sample selection. In order to see whether the dependence
between various parameters can be reduced, we focused on the more
correlated parameters reported in~\cite{2014ApJ...783..130R}, i.e.,
$r_{(R_0,\Theta_0)}$, $r_{(\Theta_0,V_{\sun})}$, $r_{(\Theta_0,\bar{V}_s)}$,
$r_{(U_{\sun},\bar{U}_s)}$ and $r_{(V_{\sun}-\bar{V}_s)}$. Figure~\ref{pearson_evol}
shows the Pearson coefficient evolution among these parameters in
sets A and B. Moreover, the Pearson coefficients calculated for the
complete sample and those published by~\cite{2014ApJ...783..130R}
are shown for comparison.                                               
 
\section{Discussion} 
\label{discussion} 
 
\subsection{Luminosity function of 6.7 GHz methanol masers} 
 
We found that $N=1300 \pm 60$ and $\alpha=-1.43 \pm 0.18$  are the
initial parameters that best match the MMB results. Using these values,
the number of sources detected and the slope of the flux density
function ($\beta$) are slightly underestimated with respect to the
observational survey results (see Table~\ref{limitations}). This
difference could be related to the contamination from inner Galaxy
sources included in the MMB, which were not included in the simulation.
This can account for approximately 100 sources in the $N$ estimate,
producing a value of $N=1300_{-160}^{+60}$.  This estimate
seems to be consistent with the initial calculation made by~\cite{2011MNRAS.417.2500G}
of $N=1250$ and also with the minimum value settled by~\cite{pan_thesis}
of $N=1125$. Moreover,~\cite{2011MNRAS.417.2500G} reported $\alpha=-1.44\pm0.4$
using kinematic distance resolution data from the International Galactic
Plane Survey, which is very close to our calculation and also gives support to our estimate of $N$ since in our method
the two quantities were fitted simultaneously.                             
 
There is no physical argument that predicts the luminosity function
to be a single power law distribution.  However, for the scope of
this paper, we are only interested in deriving an empirical relation
for the peak luminosity function for a population of 6.7 GHz methanol
masers with the proper characteristics. Additionally, a single
power law peak luminosity function appears to be consistent with
the results obtained for different systematic surveys (including
the Arecibo survey, see Figure~\ref{fluxdensityMMBAr}) and, for bright
sources, it has been previously suggested by several authors~\citep[e.g.][]{2007ApJ...656..255P,
2011MNRAS.417.2500G}.                                                  
 
\subsection{Galactic parameters analysis} 
\label{GP_analysis} 
 
The different samples described in Sect.~\ref{GP_Results} were created
to test how accurately the BeSSeL methodology can determine the Galactic
parameters. When the sample testing was initially made using the
same fitting procedure employed in~\cite{2009ApJ...700..137R,2014ApJ...783..130R},
the resulting parameters start deviating from the initial parameters
when more sources were added. When sources with large fractional errors
in parallax are numerous, we found that this biases the determination towards larger distances, resulting
in parameters that map to a bigger Galaxy. This observational effect~\citep[see
e.g.][]{2015PASP..127..994B} was corrected by allowing the fitting
procedure to de-bias distance estimations based on measured parallax.
We note that the improvements to the fitting code do not alter the
results in~\cite{2014ApJ...783..130R}, which was based on the brighter
sources.  
 
Figures~\ref{eight} and~\ref{plot_moresources_evolution} summarize
the Galactic parameters obtained compared with the initial values
adopted (see Table~\ref{initial}), using the current and possible
future samples. The results in Table~\ref{Model_BeSSeL} and Fig.~\ref{eight}
obtained for 100 simulated galaxies using the BeSSeL data sample
selection show that the Galactic parameter values can be determined
very robustly. Figure~\ref{plot_moresources_evolution} shows that
the Galactic parameter results for the simulated samples of 100 sources
(current BeSSeL data) in sets A and B are already very close to the
initial parameters, and as more sources are added the uncertainties
become smaller.                                                         
 
\subsubsection{Fundamental Galactic parameters: $R_0$, $\Theta_0$,
and $d\Theta/dR$}                                                       
 
The differences in $R_0$ and $\Theta_0$ found when using 100 simulated
galaxies mimicking the BeSSeL sample selection (Table~\ref{Model_BeSSeL}
and Fig.~\ref{eight}) are less than 0.2$\%$, demonstrating that
indeed we can recover these parameters from the adopted model, even
with samples that only contain northern hemisphere sources. Furthermore,
the errors reported by~\cite{2014ApJ...783..130R} for these parameters
(i.e. 0.16 kpc for $R_0$ and 8 $\rm{km \, s^{-1}}$ for $\Theta_0$),
which are represented in Fig.~\ref{eight} as gray regions, are double
compared to our findings. Consequently, we conclude that the errors
assigned by~\cite{2014ApJ...783..130R} to $R_0$ and $\Theta_0$ are
conservative, and there do not appear to be any bias, given the available
maser samples so far.                                                   
 
For the rotation curve, the situation is somewhat different. Although
the values found for $d\Theta/dR$ are very close to the initial values
adopted, the statistical spread is larger than expected. \cite{2014ApJ...783..130R}
reported an error of 0.4 $\rm{km \, s^{-1} \, kpc^{-1}}$ in the rotation
curve which is optimistic compared with our findings. From our simulations,
we would constrain the rotation curve value as -0.3 $\pm$ 0.9 $\rm{km
\, s^{-1} \, kpc^{-1}}$ given the BeSSeL data sample selection. The
larger error possibly indicates that our assumed velocity distributions
are too wide.                                                          
 
\begin{figure*}%
  \centering 
  \resizebox{\hsize}{!}{ \includegraphics{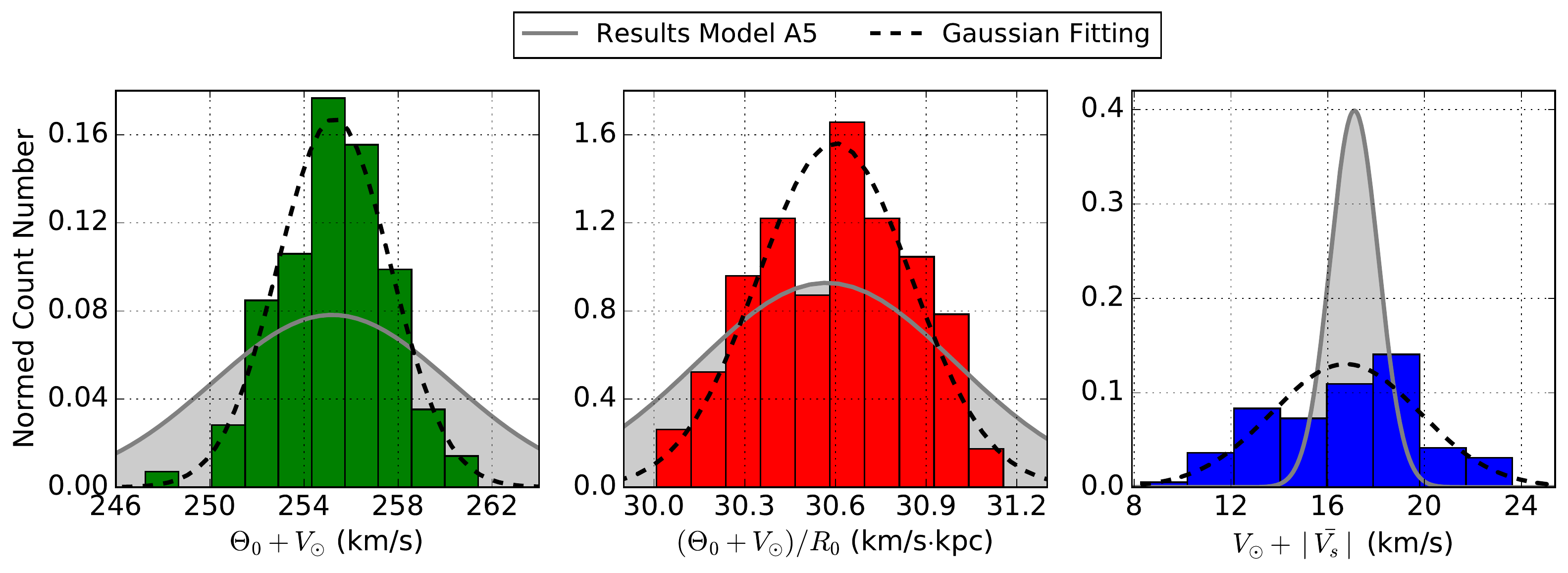}}
  \caption{\label{PDFs} Marginalized posterior probability density
distributions for correlated circular velocity parameters from 100
simulated galaxies mimicking the BeSSeL sample. Left panel: Circular
orbital speed of the Sun. Middle panel: Orbital angular solar
speed. Right panel: Difference between the circular solar motion
and average source peculiar motions.}                                   
\end{figure*}%
 
When more sources are added to the sample selection (sets A and B),
Figure~\ref{plot_moresources_evolution} shows an initial improvement
in the accuracy of the fundamental Galactic parameters. For set A,
the errors in $R_0$, $\Theta_0$, and $d\Theta/dR$ can improve up to
$\pm \, 0.04 \, \rm{kpc}$, $\pm \, 2.7 \, \rm{km \, s^{-1}}$, and
$\pm \, 0.6 \, \rm{km \, s^{-1} \, kpc^{-1}}$, respectively, when
more northern hemisphere sources are added. In contrast when southern
hemisphere sources are also added (set B), the errors can decrease
to $\pm \, 0.03 \, \rm{kpc}$, $\pm \, 2.5 \, \rm{km \, s^{-1}}$, and
$\pm \, 0.3 \, \rm{km \, s^{-1} \, kpc^{-1}}$, respectively. Further
improvements would require much better astrometry for weak sources,
requiring for example much more sensitive observations.               
 
\subsubsection{Solar motion and average source peculiar motions} 
 
Figure~\ref{eight} shows the distribution obtained for solar velocity
components ($U_{\sun}$, $V_{\sun}$, and $W_{\sun}$). The uncertainties
derived were around 2$\%$ for all solar velocity components with
respect to the initial parameters. For $U_{\sun}$, the standard deviation
found was 1.2 $\rm{km \, s^{-1}}$, which compares well to the results
published by~\cite{2014ApJ...783..130R} (i.e., $U_{\sun}= \pm 1.8
\, \rm{km \, s^{-1}}$). In addition, the spread for $V_{\sun}$ and
$W_{\sun}$ are narrower with values of 2.4 $\rm{km \, s^{-1}}$ and
0.5 $\rm{km \, s^{-1}}$, respectively. Compared to the BeSSeL results
(i.e., $V_{\sun} = \pm 6.8 \, \rm{km \, s^{-1}}$ and $W_{\sun}= \pm
2.1 \, \rm{km \, s^{-1}}$), we can affirm that the solar motion results
published by~\cite{2014ApJ...783..130R} have conservative estimates.    
 
For the radial and tangential average peculiar motions ($\bar{U}_s$
and $\bar{V}_s$), Table~\ref{Model_BeSSeL} shows that indeed these
peculiar velocities can be fitted with high accuracy using the simulations;
however, the relative spreads are high compared with other parameters
(see Figure~\ref{eight}). Even so, compared to the BeSSeL
results (i.e., $\bar{U}_s= \pm 2.1 \, \rm{km \, s^{-1}}$ and $\bar{V}_s=
\pm 6.8 \, \rm{km \, s^{-1}}$), the errors in the simulated sample
selection are still lower (i.e., $\bar{U}_s= \pm 1.7 \, \rm{km \,
s^{-1}}$ and $\bar{V}_s= \pm 3.8 \, \rm{km \, s^{-1}}$).                
 
The reason for higher dispersions for $\bar{U}_s$ and $\bar{V}_s$
could be related to the number of parameters that must fit independently. The parameters $\bar{U}_s$ and $U_{\sun}$ have largely the same effect on the observations
for nearby sources and therefore are directly correlated (see Table~\ref{pearsons}).
For $\bar{V}_s$, the correlation is high with two components ($V_{\sun}$,
$\Theta_0$), which affects the fitting and hence the estimated accuracy.
 
\subsubsection{Parameter correlations} 
\label{correlation} 
 
Pearson product-moment correlation coefficients are shown in Table~\ref{pearsons}.
All the parameters are in agreement with those found by~\cite{2014ApJ...783..130R},
except for the Pearson coefficient between $d\Theta/dR$ and $\bar{V}_s$
(labeled "$r_{d\Theta/dR,\bar{V}_s}$") where a low correlation was
found in the observed sample instead of the moderate correlation
found in the simulated sample. However, we focus the
discussion on the parameters that were reported to have considerable
correlation in~\cite{2014ApJ...783..130R}, as it is interesting to
see how it is possible to disentangle these fundamental parameters.               
 
For the first BeSSeL summary paper~\citep{2009ApJ...700..137R},
where only 16 HMSFRs were used, the estimated $R_0$ and $\Theta_0$
were strongly correlated ($r_{R_0,\Theta_0}=0.87$). Later on, in~\cite{2014ApJ...783..130R},
using a larger sample of 103 HMSFRs spanning a greater Galactic 
distribution, the correlation was significantly lower ($r_{R_0,\Theta_0}=0.47$).
We calculated the same coefficient by mimicking both samples using
100 simulated galaxies, finding similar results for each sample,
i.e,  $r_{R_0,\Theta_0}=0.77$ and $r_{R_0,\Theta_0}=0.48$. These
results show that, indeed, our simulation produces similar correlation
coefficients to those found in the observations, even when the method used to
calculate the Pearson coefficients are completely different (see
Sect.~\ref{correlations}). When more sources are added to the sample
selection, Figure~\ref{pearson_evol} shows that the correlation between
$R_0$ and $\Theta_0$ is reduced. Furthermore, the Pearson coefficient
will have a moderate value (0.3) when using the complete data sample,
which demonstrates that the correlation between these parameters
can be unraveled smoothly as more sources are added.                    
 
For the tangential velocity component, we have three different Galactic
parameters giving similar effects: $\Theta_0$, $V_{\sun}$ and $\bar{V}_s$.
Figure~\ref{pearson_evol} and Table~\ref{pearsons} show that the
Pearson coefficients among these parameters are always high, even
when more sources are added. This implies that different types of 
data are needed in order to better disentangle these Galactic parameters. Finally,
in the radial direction we have two Galactic parameters: $U_{\sun}$
and $\bar{U}_s$. The correlation between these parameters is around
0.5 for a low number of sources, as it shown in Figure~\ref{pearson_evol}
and listed in Table~\ref{pearsons}. When more sources are added,
the correlation parameter seems to maintain the same value or slightly
increase in both sets of samples (see Figure~\ref{pearson_evol}).
One could expect that the inclusion of southern hemisphere sources
would help to disentangle some of the dependences; however, comparing
the top and bottom plots in Figure~\ref{pearson_evol} we can see
that using samples with a larger coverage of the Galaxy does not alter
significantly the correlation values found for any of the Pearson
coefficients discussed here.                                            
 
As some of the parameter correlations persist, even when more sources are added
to the sample selection, we estimated the marginalized PDFs for different
combined parameters that were also reported in the latest BeSSeL survey paper.
Figure~\ref{PDFs} shows the PDFs for the circular orbital speed of
the Sun, the orbital angular solar speed, and the difference between the
circular solar and average source peculiar motions. Additionally,
those PDFs reported by~\cite{2014ApJ...783..130R}
are shown as gray regions.                                              
 
We found $255.3 \pm 2.4 \, \rm{km \, s^{-1}}$, $30.60 \pm 0.26 \,
\rm{km \, s^{-1} \, kpc^{-1}}$, and $16.7 \pm 3.1 \, \rm{km \, s^{-1}}$,
respectively, for the correlated values of the parameters $\Theta_0
+ V_{\sun}$, $(\Theta_0 + V_{\sun})/R_0$ and $V_{\sun}+\mid \bar{V}_s
\mid$. \cite{2014ApJ...783..130R} found more conservative values for $\Theta_0
+ V_{\sun}=  255.2 \pm 5.1 \, \rm{km \, s^{-1}} $ and $(\Theta_0
+ V_{\sun})/R_0= 30.57 \pm 0.43 \, \rm{km \, s^{-1}kpc^{-1}} $ than we did. Although the mean value found for $V_{\sun}+ \mid
\bar{V}_s \mid$ is in agreement with the BeSSeL results (i.e., $17.1
\pm 1.0 \, \rm{km \, s^{-1}}$), we estimate a wider error of $\pm
3.1 \, \rm{km \, s^{-1}}$ based on our simulations.                        
 
 
\section{Conclusions} 
\label{conclusions}

We constructed simulations of 6.7 GHz methanol maser
distributions to test whether the Galactic parameter results obtained
by the BeSSeL survey~\citep{2014ApJ...783..130R} are biased in any
way and investigated the interdependencies between some parameter
estimates.  We used our model to constrain the peak flux density
function for the masers and obtained similar results to those of
systematic unbiased surveys (MMB and Arecibo).
This comparison allowed us to estimate the integral number
of sources ($N=1300_{-160}^{+60}$) and the slope of the luminosity
function ($\alpha=-1.43\pm0.18$), which showed good agreement with~\cite{pan_thesis,2011MNRAS.417.2500G,2013MNRAS.431.1752U}.

Assuming that the observations are predominantly of 12
GHz methanol masers found through 6.7 GHz surveys, we simulated the
current database of the BeSSeL survey hundreds of times. We found
that the fundamental Galactic parameters ($R_0$, $\Theta_0$, $d\Theta/dR$),
the solar velocity components ($U_{\sun}$,$V_{\sun}$, $W_{\sun}$)
and the average peculiar motion ($\bar{U}_s$, $\bar{V}_s$) can be
determined robustly. Furthermore, the results published by~\cite{2014ApJ...783..130R}
have a conservative error calculation given the current sample, except possibly
for the rotation curve error estimate. Also, correlation coefficients
for the various Galactic parameters in our simulations and those reported by~\cite{2014ApJ...783..130R} are similar.        
 
Additionally, the fitting procedures developed by~\cite{2009ApJ...700..137R,2014ApJ...783..130R}
for use with the BeSSeL data and improved in this study estimate
Galactic parameters correctly even when weak and/or distant sources 
with large fractional parallax uncertainties are included in the samples. 
                                     
For future BeSSeL observations, the simulations demonstrate that
the Galactic parameter estimates can be improved and the error bars
reduced significantly. Moreover, using southern hemisphere data,
the Galactic parameter estimates improve notably compared with samples
limited to the northern sky.                                            
 
We find that the uncertainties in the values of certain combined velocity parameters 
that are highly correlated are similar to those published in~\cite{2014ApJ...783..130R}, 
except for the dispersion in $V_{\sun}+ \mid \bar{V}_s \mid$.  
However, when more sources are added to the sample, the correlations among most
Galactic parameters are smoothly reduced; for the highly velocity parameters the correlation coefficients do not decrease significantly.                  
 
The framework proposed to test the results of the BeSSeL survey is
useful for defining requirements for future astrometric campaigns
that are similar or complementary to the BeSSeL data. Southern arrays
-- like the Australian Long Baseline Array~\citep[see e.g.][]{2015ApJ...805..129K,2017MNRAS.465.1095K}
and, in the future, the African VLBI Network and the Square Kilometre
Array in Australia and South Africa -- will supplement the lack of
precise astrometric data in quadrants III and IV of the Milky Way
plane, where only a few sources have been measured. Moreover, astrometric
studies that include the inner Galactic region, such as the Bulge
Asymmetries and Dynamic Evolution (BAaDE\footnote{\url{http://www.phys.unm.edu/~baade/index.html}})
project, aim to resolve the dynamics of the bar by measuring
proper motions and distances of SiO masers present in AGB stars~\citep{2017IAUS..322..103S}.
Out of the Galactic plane, Gaia will soon provide astrometric results
for a large number of sources. All of these investigations will contribute
to the determination of Galactic parameters with even better accuracy
with new and improved astrometric data. Until then, Galactic
simulations complement the current observations by demonstrating
their robustness and potential.                                         
 
 
\begin{acknowledgements} 
We sincerely thank the anonymous referee for making valuable suggestions
that have improved the paper. L.H.Q.-N. acknowledges the comments and
suggestions regarding the model implementation made by S. Solorzano-Rocha
at ETH Z\"urich.                                                        
\end{acknowledgements} 
 
 
\bibliographystyle{aa} 
\bibliography{quiroganunez_v14.bib} 

\begin{thebibliography}{37}
\expandafter\ifx\csname natexlab\endcsname\relax\def\natexlab#1{#1}\fi

\bibitem[{{Bailer-Jones}(2015)}]{2015PASP..127..994B}
{Bailer-Jones}, C.~A.~L. 2015, \pasp, 127, 994

\bibitem[{{Bobylev} \& {Bajkova}(2016)}]{2016AstL...42..182B}
{Bobylev}, V.~V. \& {Bajkova}, A.~T. 2016, Astronomy Letters, 42, 182

\bibitem[{{Bovy} \& {Rix}(2013)}]{2013ApJ...779..115B}
{Bovy}, J. \& {Rix}, H.-W. 2013, \apj, 779, 115

\bibitem[{{Breen} {et~al.}(2013){Breen}, {Ellingsen}, {Contreras}, {Green},
  {Caswell}, {Stevens}, {Dawson}, \& {Voronkov}}]{2013MNRAS.435..524B}
{Breen}, S.~L., {Ellingsen}, S.~P., {Contreras}, Y., {et~al.} 2013, \mnras,
  435, 524

\bibitem[{{Brunthaler} {et~al.}(2011){Brunthaler}, {Reid}, {Menten}, {Zheng},
  {Bartkiewicz}, {Choi}, {Dame}, {Hachisuka}, {Immer}, {Moellenbrock},
  {Moscadelli}, {Rygl}, {Sanna}, {Sato}, {Wu}, {Xu}, \&
  {Zhang}}]{2011AN....332..461B}
{Brunthaler}, A., {Reid}, M.~J., {Menten}, K.~M., {et~al.} 2011, Astronomische
  Nachrichten, 332, 461

\bibitem[{{Burns} {et~al.}(2017){Burns}, {Handa}, {Imai}, {Nagayama},
  {Omodaka}, {Hirota}, {Motogi}, {van Langevelde}, \&
  {Baan}}]{2017MNRAS.tmp..217B}
{Burns}, R.~A., {Handa}, T., {Imai}, H., {et~al.} 2017, \mnras

\bibitem[{{Caswell} {et~al.}(2010){Caswell}, {Fuller}, {Green}, {Avison},
  {Breen}, {Brooks}, {Burton}, {Chrysostomou}, {Cox}, {Diamond}, {Ellingsen},
  {Gray}, {Hoare}, {Masheder}, {McClure-Griffiths}, {Pestalozzi}, {Phillips},
  {Quinn}, {Thompson}, {Voronkov}, {Walsh}, {Ward-Thompson}, {Wong-McSweeney},
  {Yates}, \& {Cohen}}]{2010MNRAS.404.1029C}
{Caswell}, J.~L., {Fuller}, G.~A., {Green}, J.~A., {et~al.} 2010, \mnras, 404,
  1029

\bibitem[{{Caswell} {et~al.}(2011){Caswell}, {Fuller}, {Green}, {Avison},
  {Breen}, {Ellingsen}, {Gray}, {Pestalozzi}, {Quinn}, {Thompson}, \&
  {Voronkov}}]{2011MNRAS.417.1964C}
{Caswell}, J.~L., {Fuller}, G.~A., {Green}, J.~A., {et~al.} 2011, \mnras, 417,
  1964

\bibitem[{{Cheng} {et~al.}(2012){Cheng}, {Rockosi}, {Morrison}, {Lee}, {Beers},
  {Bizyaev}, {Harding}, {Malanushenko}, {Malanushenko}, {Oravetz}, {Pan},
  {Schlesinger}, {Schneider}, {Simmons}, \& {Weaver}}]{2012ApJ...752...51C}
{Cheng}, J.~Y., {Rockosi}, C.~M., {Morrison}, H.~L., {et~al.} 2012, \apj, 752,
  51

\bibitem[{{Dame} {et~al.}(2001){Dame}, {Hartmann}, \&
  {Thaddeus}}]{2001ApJ...547..792D}
{Dame}, T.~M., {Hartmann}, D., \& {Thaddeus}, P. 2001, \apj, 547, 792

\bibitem[{{Efremov}(2011)}]{2011ARep...55..108E}
{Efremov}, Y.~N. 2011, Astronomy Reports, 55, 108

\bibitem[{{Gaia Collaboration} {et~al.}(2016){Gaia Collaboration}, {Brown},
  {Vallenari}, {Prusti}, {de Bruijne}, {Mignard}, {Drimmel}, {Babusiaux},
  {Bailer-Jones}, {Bastian}, \& et~al.}]{2016A&A...595A...2G}
{Gaia Collaboration}, {Brown}, A.~G.~A., {Vallenari}, A., {et~al.} 2016, \aap,
  595, A2

\bibitem[{{Gilmore} \& {Reid}(1983)}]{1983MNRAS.202.1025G}
{Gilmore}, G. \& {Reid}, N. 1983, \mnras, 202, 1025

\bibitem[{{Goodman} {et~al.}(2014){Goodman}, {Alves}, {Beaumont}, {Benjamin},
  {Borkin}, {Burkert}, {Dame}, {Jackson}, {Kauffmann}, {Robitaille}, \&
  {Smith}}]{2014ApJ...797...53G}
{Goodman}, A.~A., {Alves}, J., {Beaumont}, C.~N., {et~al.} 2014, \apj, 797, 53

\bibitem[{{Green} {et~al.}(2009){Green}, {Caswell}, {Fuller}, {Avison},
  {Breen}, {Brooks}, {Burton}, {Chrysostomou}, {Cox}, {Diamond}, {Ellingsen},
  {Gray}, {Hoare}, {Masheder}, {McClure-Griffiths}, {Pestalozzi}, {Phillips},
  {Quinn}, {Thompson}, {Voronkov}, {Walsh}, {Ward-Thompson}, {Wong-McSweeney},
  {Yates}, \& {Cohen}}]{2009MNRAS.392..783G}
{Green}, J.~A., {Caswell}, J.~L., {Fuller}, G.~A., {et~al.} 2009, \mnras, 392,
  783

\bibitem[{{Green} {et~al.}(2010){Green}, {Caswell}, {Fuller}, {Avison},
  {Breen}, {Ellingsen}, {Gray}, {Pestalozzi}, {Quinn}, {Thompson}, \&
  {Voronkov}}]{2010MNRAS.409..913G}
{Green}, J.~A., {Caswell}, J.~L., {Fuller}, G.~A., {et~al.} 2010, \mnras, 409,
  913

\bibitem[{{Green} {et~al.}(2012){Green}, {Caswell}, {Fuller}, {Avison},
  {Breen}, {Ellingsen}, {Gray}, {Pestalozzi}, {Quinn}, {Thompson}, \&
  {Voronkov}}]{2012MNRAS.420.3108G}
{Green}, J.~A., {Caswell}, J.~L., {Fuller}, G.~A., {et~al.} 2012, \mnras, 420,
  3108

\bibitem[{{Green} {et~al.}(2011){Green}, {Caswell}, {McClure-Griffiths},
  {Avison}, {Breen}, {Burton}, {Ellingsen}, {Fuller}, {Gray}, {Pestalozzi},
  {Thompson}, \& {Voronkov}}]{2011ApJ...733...27G}
{Green}, J.~A., {Caswell}, J.~L., {McClure-Griffiths}, N.~M., {et~al.} 2011,
  \apj, 733, 27

\bibitem[{{Green} \& {McClure-Griffiths}(2011)}]{2011MNRAS.417.2500G}
{Green}, J.~A. \& {McClure-Griffiths}, N.~M. 2011, \mnras, 417, 2500

\bibitem[{{Krishnan} {et~al.}(2017){Krishnan}, {Ellingsen}, {Reid}, {Bignall},
  {McCallum}, {Phillips}, {Reynolds}, \& {Stevens}}]{2017MNRAS.465.1095K}
{Krishnan}, V., {Ellingsen}, S.~P., {Reid}, M.~J., {et~al.} 2017, \mnras, 465,
  1095

\bibitem[{{Krishnan} {et~al.}(2015){Krishnan}, {Ellingsen}, {Reid},
  {Brunthaler}, {Sanna}, {McCallum}, {Reynolds}, {Bignall}, {Phillips},
  {Dodson}, {Rioja}, {Caswell}, {Chen}, {Dawson}, {Fujisawa}, {Goedhart},
  {Green}, {Hachisuka}, {Honma}, {Menten}, {Shen}, {Voronkov}, {Walsh}, {Xu},
  {Zhang}, \& {Zheng}}]{2015ApJ...805..129K}
{Krishnan}, V., {Ellingsen}, S.~P., {Reid}, M.~J., {et~al.} 2015, \apj, 805,
  129

\bibitem[{{Pandian}(2007)}]{pan_thesis}
{Pandian}, J.~D. 2007, PhD thesis, Cornell University

\bibitem[{{Pandian} {et~al.}(2007){Pandian}, {Goldsmith}, \&
  {Deshpande}}]{2007ApJ...656..255P}
{Pandian}, J.~D., {Goldsmith}, P.~F., \& {Deshpande}, A.~A. 2007, \apj, 656,
  255

\bibitem[{{Perryman} {et~al.}(2001){Perryman}, {de Boer}, {Gilmore}, {H{\o}g},
  {Lattanzi}, {Lindegren}, {Luri}, {Mignard}, {Pace}, \& {de
  Zeeuw}}]{2001A&A...369..339P}
{Perryman}, M.~A.~C., {de Boer}, K.~S., {Gilmore}, G., {et~al.} 2001, \aap,
  369, 339

\bibitem[{{Perryman} {et~al.}(1997){Perryman}, {Lindegren}, {Kovalevsky},
  {Hoeg}, {Bastian}, {Bernacca}, {Cr{\'e}z{\'e}}, {Donati}, {Grenon},
  {Grewing}, {van Leeuwen}, {van der Marel}, {Mignard}, {Murray}, {Le Poole},
  {Schrijver}, {Turon}, {Arenou}, {Froeschl{\'e}}, \&
  {Petersen}}]{1997A&A...323L..49P}
{Perryman}, M.~A.~C., {Lindegren}, L., {Kovalevsky}, J., {et~al.} 1997, \aap,
  323, L49

\bibitem[{{Pestalozzi} {et~al.}(2007){Pestalozzi}, {Chrysostomou}, {Collett},
  {Minier}, {Conway}, \& {Booth}}]{2007A&A...463.1009P}
{Pestalozzi}, M.~R., {Chrysostomou}, A., {Collett}, J.~L., {et~al.} 2007, \aap,
  463, 1009

\bibitem[{{Reid} \& {Honma}(2014)}]{2014ARA&A..52..339R}
{Reid}, M.~J. \& {Honma}, M. 2014, \araa, 52, 339

\bibitem[{{Reid} {et~al.}(2014){Reid}, {Menten}, {Brunthaler}, {Zheng}, {Dame},
  {Xu}, {Wu}, {Zhang}, {Sanna}, {Sato}, {Hachisuka}, {Choi}, {Immer},
  {Moscadelli}, {Rygl}, \& {Bartkiewicz}}]{2014ApJ...783..130R}
{Reid}, M.~J., {Menten}, K.~M., {Brunthaler}, A., {et~al.} 2014, \apj, 783, 130

\bibitem[{{Reid} {et~al.}(2009){Reid}, {Menten}, {Zheng}, {Brunthaler},
  {Moscadelli}, {Xu}, {Zhang}, {Sato}, {Honma}, {Hirota}, {Hachisuka}, {Choi},
  {Moellenbrock}, \& {Bartkiewicz}}]{2009ApJ...700..137R}
{Reid}, M.~J., {Menten}, K.~M., {Zheng}, X.~W., {et~al.} 2009, \apj, 700, 137

\bibitem[{{Rix} \& {Bovy}(2013)}]{2013A&ARv..21...61R}
{Rix}, H.-W. \& {Bovy}, J. 2013, \aapr, 21, 61

\bibitem[{{Sanna} {et~al.}(2014){Sanna}, {Reid}, {Menten}, {Dame}, {Zhang},
  {Sato}, {Brunthaler}, {Moscadelli}, \& {Immer}}]{2014ApJ...781..108S}
{Sanna}, A., {Reid}, M.~J., {Menten}, K.~M., {et~al.} 2014, \apj, 781, 108

\bibitem[{{Sjouwerman} {et~al.}(2017){Sjouwerman}, {Pihlstr{\"o}m}, {Rich},
  {Morris}, \& {Claussen}}]{2017IAUS..322..103S}
{Sjouwerman}, L.~O., {Pihlstr{\"o}m}, Y.~M., {Rich}, R.~M., {Morris}, M.~R., \&
  {Claussen}, M.~J. 2017, in IAU Symposium, Vol. 322, The Multi-Messenger
  Astrophysics of the Galactic Centre, ed. R.~M. {Crocker}, S.~N. {Longmore},
  \& G.~V. {Bicknell}, 103--106

\bibitem[{{Surcis} {et~al.}(2013){Surcis}, {Vlemmings}, {van Langevelde},
  {Hutawarakorn Kramer}, \& {Quiroga-Nu{\~n}ez}}]{2013A&A...556A..73S}
{Surcis}, G., {Vlemmings}, W.~H.~T., {van Langevelde}, H.~J., {Hutawarakorn
  Kramer}, B., \& {Quiroga-Nu{\~n}ez}, L.~H. 2013, \aap, 556, A73

\bibitem[{{Urquhart} {et~al.}(2013){Urquhart}, {Moore}, {Schuller}, {Wyrowski},
  {Menten}, {Thompson}, {Csengeri}, {Walmsley}, {Bronfman}, \&
  {K{\"o}nig}}]{2013MNRAS.431.1752U}
{Urquhart}, J.~S., {Moore}, T.~J.~T., {Schuller}, F., {et~al.} 2013, \mnras,
  431, 1752

\bibitem[{{van der Walt}(2005)}]{2005MNRAS.360..153V}
{van der Walt}, J. 2005, \mnras, 360, 153

\bibitem[{{Wainscoat} {et~al.}(1992){Wainscoat}, {Cohen}, {Volk}, {Walker}, \&
  {Schwartz}}]{1992ApJS...83..111W}
{Wainscoat}, R.~J., {Cohen}, M., {Volk}, K., {Walker}, H.~J., \& {Schwartz},
  D.~E. 1992, \apjs, 83, 111

\bibitem[{{Yusof} {et~al.}(2013){Yusof}, {Hirschi}, {Meynet}, {Crowther},
  {Ekstr{\"o}m}, {Frischknecht}, {Georgy}, {Abu Kassim}, \&
  {Schnurr}}]{2013MNRAS.433.1114Y}
{Yusof}, N., {Hirschi}, R., {Meynet}, G., {et~al.} 2013, \mnras, 433, 1114

\end{thebibliography}
 
\end{document}